\documentstyle[graphicx]{mn2e}
\newif\ifAMStwofonts

\title{The spatial distribution of cold gas in hierarchical galaxy formation models}
\author[Han-Seek Kim et al.]
       {Han-Seek~Kim$^1$\thanks{h.s.kim@durham.ac.uk}, C.M.~Baugh$^1$,  A.J.~Benson$^2$, S.~Cole$^1$, C.S.~Frenk$^1$,C.G.~Lacey$^1$, \vspace{0.3cm} \\ \hspace{-0.1cm}\LARGE{ C. Power $^{3}$, M.~Schneider$^1$} \\
       $^1$Institute for Computational Cosmology, Department of Physics, University of Durham, South Road, Durham DH1 3LE, UK\\
       $^2$Theoretical Astrophysics, Caltech, MC350-17, 1200 E. California Blvd., Pasadena CA 91125, USA\\
       $^3$Department of Physics and Astronomy, University of Leicester, Leicester LE1 7RH, UK}
\date{}

\pagerange{\pageref{firstpage}--\pageref{lastpage}}
\pubyear{2009}

\begin{document}

\maketitle
\title{Where is the cold gas?}
\label{firstpage}

\begin{abstract}
The distribution of cold gas in dark matter haloes is driven by key processes in galaxy formation: 
gas cooling, galaxy mergers, star formation and reheating of gas by supernovae. We compare the predictions 
of four different galaxy formation models for the spatial distribution of cold gas. We find that satellite 
galaxies make little contribution to the abundance or clustering strength of cold gas selected samples, and 
are far less important than they are in optically selected samples. The halo occupation distribution 
function of present-day central galaxies with cold gas mass $> 10^{9} h^{-1} M_{\odot}$ is peaked 
around a halo mass of $\approx 10^{11} h^{-1} M_{\odot}$, a scale that is set by the AGN suppression 
of gas cooling. The model predictions for the projected correlation function 
are in good agreement with measurements from the HI Parkes All-Sky Survey. 
We compare the effective volume of possible surveys with the Square Kilometre 
Array\footnote{http://www.skatelescope.org} with those expected for a redshift 
survey in the near-infrared. Future redshift surveys using neutral hydrogen emission 
will be competitive with the most ambitious spectroscopic surveys planned in the near-infrared.
\end{abstract}


\section{Introduction}

Cold gas is central to galaxy formation yet little is known about how much there is in the 
Universe at different epochs and how this gas is distributed in dark matter haloes of different mass. 
The primary probe of atomic hydrogen, 21cm line emission, is incredibly weak. It is only in recent years 
that a robust and comprehensive census of atomic hydrogen (HI) in the local universe has been made possible through 
the HI Parkes All Sky Survey (Barnes et~al. 2001; Zwaan et~al. 2003, 2005). This work is being 
extended to lower mass systems by the ALFALFA survey (Giovanelli et~al. 2007). 
Despite this progress, the highest redshift direct detection of HI in emission is very firmly confined to the 
local Universe at $z=0.34$ (Lah et~al. 2009, see also Verheijen et~al. 2007). 
Information about cold gas in the high redshift Universe is restricted to  
absorption lines in quasar spectra (e.g. Peroux et~al. 2003). However, over the coming decade, 
this situation is expected to change dramatically with the construction of new, more sensitive 
radio telescopes such as the pathfinders for the Square Kilometre Array, MeerKAT (Booth et~al. 2009) 
and ASKAP (Johnston et~al. 2008), and the Square Kilometre Array itself (Schilizzi, Dewdney \& Lazio 2008). 
The SKA will revolutionise our understanding of galaxy formation and cosmology, uncovering the HI Universe 
out to high redshifts. One of the major science goals is to better characterise the evolution of 
dark energy with redshift. The SKA is expected to provide competitive constraints on 
the nature of dark energy through high accuracy measurement of  large-scale structure in the 
galaxy distribution over a lookback time representing a significant fraction of the age of the 
Universe (Albrecht et~al. 2006). This conclusion currently rests on very uncertain calculations 
which we seek to place on a firmer, more physical footing in this paper. 

Modelling the abundance and clustering of HI sources is challenging. 
A number of possible approaches have been tried; empirical modelling, which relies upon the 
observations of HI in the Universe, the fully numerical approach, which uses cosmological 
gas dynamics simulations to model the HI content of galaxies from first principles and semi-analytical 
modelling, which we use in this paper. Empirical estimates have been attempted despite the paucity of 
observational results for guidance (Abdalla \& Rawlings 2005; Abdalla, Blake \& Rawlings 2010). 
Such calculations require an assumption about the evolution of the HI mass function over a broad 
redshift interval. The only constraint on this assumption is the integrated density of HI, which can be 
compared with the results inferred from quasar absorption features, which themselves require corrections 
for unseen low column density systems and dust extinction (Storrie-Lombardi et~al. 1996). The empirical 
approach does not predict the clustering of HI sources. Further assumptions and approximations 
are necessary to extend this class of modelling so that predictions can be made for galaxy clustering. 
Another layer of approximation in this class of modelling 
has been motivated by observations which suggest that HI sources tend to avoid the centres of 
clusters and that clusters do not boast an important population of satellites 
(e.g. Waugh et~al. 2002; Verheijen et~al. 2007). This led Marin et~al. (2009) to make a one-to-one 
connection between halo mass and HI mass. However, the nature of the relation is uncertain and 
several possibilities are explored by Marin et~al. based on different assumptions about the 
evolution of the HI mass function. 

Ideally, a physically motivated model which follows the sources and sinks of cold gas is needed. 
Gas dynamic simulations are computationally expensive and are typically restricted 
to small computational volumes, which makes it impossible to accurately follow the growth of 
structure to the present day. An example is provided by Popping et~al. (2009), who carry out a 
smoothed particle hydrodynamics simulation in a $32 h^{-1}$Mpc box. The HI mass function in the 
simulation is in very poor agreement with the observational estimate of Zwaan et~al. (2005), 
underpredicting the abundance of galaxies of HI mass $10^{10} M_{\odot}$ by a factor of 30, which 
the authors put down to the small computational volume, and overpredicting low mass systems by 
a factor of two. Clustering predictions are limited to scales smaller than a few Mpc due to 
the small box size. Furthermore, it is important to be aware that gas dynamic simulations do not 
have the resolution to follow all of the processes in galaxy formation directly and in all cases 
resort to what are essentially semi-analytical rules to treat sub-resolution physics. 

Currently the most promising route to making physical and robust predictions for the HI in the 
Universe is semi-analytical modelling of galaxy formation (see Baugh 2006). This type of model 
includes a simplified but physically motivated treatment of the processes which control the 
amount of cold gas in a galaxy: gas cooling, galaxy mergers, star formation and reheating of 
gas by supernovae. These calculations are quick and 
can rapidly cover the haloes in a cosmological volume. Baugh et~al. (2004) presented predictions for the mass 
function of cold gas galaxies in the {\tt GALFORM} semi-analytical model of Cole et~al. (2000). 
One issue which must be dealt with is that the models predict only the total mass of cold gas, which includes helium, 
and both atomic and molecular hydrogen. Baugh et~al. assumed a fixed ratio of molecular to atomic hydrogen. 
Obreschkow \& Rawlings (2009) developed an empirical model based on observations and theoretical arguments 
by Blitz \& Rosolowsky (2006) in which this ratio could vary from galaxy to galaxy.  Obreschkow \& Rawlings applied 
this ansatz to the semi-analytical model of de Lucia \& Blaizot (2007). 

In the first paper in this series, 
we compared the predictions of a range of semi-analytical models for the mass function 
of HI (Power et~al. 2009). Despite the different implementations of the physical ingredients used in the models 
and the different emphasis placed on various observations when setting the model parameters, the predictions show 
generic features. Power et~al. found that there is surprisingly little variation in the predicted HI mass function 
with redshift, and that the models make similar predictions for the rotation speed and size of HI systems. 
The models predict the mass of cold gas and so a conversion is required to turn this into a HI mass. Currently 
the most uncertain step is the assumption about what fraction of hydrogen is in atomic form and what fraction 
is molecular. Power et~al. presented predictions for two cases, one in which all model galaxies are assumed to have 
a fixed molecular to atomic hydrogen ratio ($H_{2}/$HI) and the other in which this ratio varies from galaxy 
to galaxy, depending upon the local conditions in the galactic disk (Blitz \& Rosolowsky 2006). 
The assumption of a variable $H_{2}/$HI ratio results in a dramatic reduction in the number of HI 
sources in the tail of the redshift distribution. 

In this paper we look at the distribution of cold gas in galaxies as a function of halo mass. 
In particular we look at the halo occupation distribution (HOD) for HI galaxies, 
which gives the mean number of galaxies of a given HI mass as a function of dark 
matter halo mass, and the clustering of HI galaxies. Using this information, we 
assess the potential of the SKA to measure the baryonic acoustic oscillation (BAO) signal.
We briefly review the {\tt GALFORM} model in Section 2, explaining the differences between the four models 
that we consider. We then look at the halo occupation distribution of cold gas galaxies in Section 3, 
in which we also present predictions for the clustering of cold gas galaxies at different redshifts 
and compare to measured clustering at the present day. In Section 4 we compare the performance of 
future redshift surveys in the optical and using HI emission for measuring the properties of the 
dark energy. We present a summary along with our conclusions in Section 5.

\section{Galaxy formation models and basic predictions} \label{model}

\begin{figure}
\includegraphics[width=8.6cm]{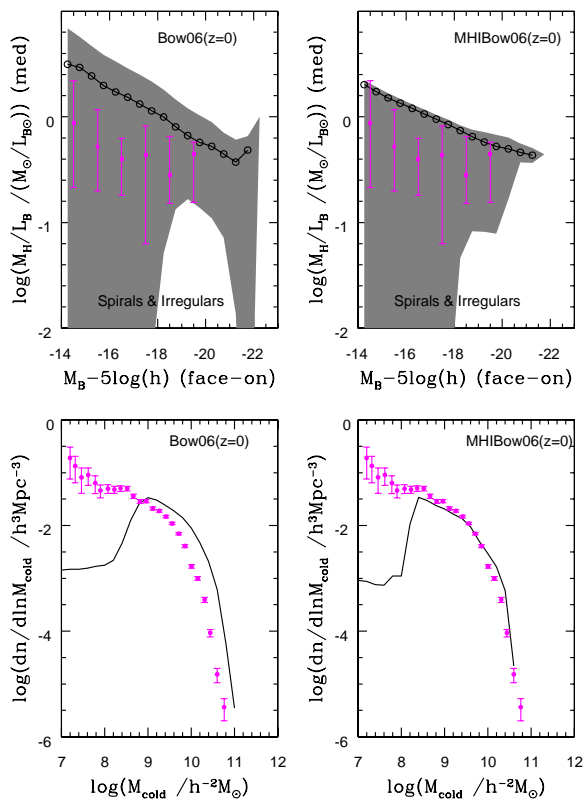}
\caption{
The predicted ratio of neutral hydrogen mass to B-band luminosity luminosity (upper panels) and 
the cold gas mass function (lower panels) in the Bow06 (left panels) and MHIBow06 models 
(right panels). In the upper panels, the magenta points show observational estimates of 
the hydrogen mass to luminosity ratio using data from Huctmeier \& Richter (1988) (HI) 
and Sage (1993) (H$_{2}$). The black points show the median ratio predicted by the models and 
the grey shading shows the 20 - 80 percentile range of the predicted distribution. 
We assume that 76\% by mass of the cold gas mass predicted by the models is neutral hydrogen. 
In the lower panels, the magenta points show the cold gas mass function derived from the HI 
mass function estimated by Zwaan et~al (2005). Here, a constant H$_{2}/$HI ratio of 0.4 has 
been assumed to convert the HI measurement into a cold gas mass. 
} 
\label{COMCOLD}
\end{figure}

\begin{table*}
\caption{
The values of selected parameters which are different in the models. 
The columns are as follows: (1) The name of the model. (2) The equation used to calculate the star formation 
timescale, $\tau_{\star}$. (3) The value of $\epsilon_{\star}$ or $\tau_{0}^\star$ used in the 
star formation timescale. (4) The AGN feedback parameter, $\alpha_{\rm{cool}}$, (Eq.~\ref{AGN}) 
(5) The supernova feedback parameter,$V_{\rm{hot}}$ (Eq.~\ref{SNe}). 
(6) The source of halo merger histories. 
(7) Comments giving model source or key differences from published models.}
\label{Parameters}
\begin{tabular}{lcccccl}
\hline
\hline
 & $\tau_{\star}$&$\epsilon_{\star}$ or $\tau^{0}_{\star}$ [Gyr]&$\alpha_{\rm{cool}}$&$V_{\rm{hot}}$[kms$^{-1}$]&
Merger & Comments\\
 &               &                                              &                    &                          & 
tree   &         \\
\hline
Bow06 & Eq.~\ref{st2}  &0.0029 & 0.58& 485  & N-body     & Bower et~al. (2006)\\
Font08& Eq.~\ref{st2}  &0.0029 & 0.70& 485  & N-body     & Font et~al. (2008) \\
      &                &       &     &      &            & Modified cooling recipe in satellites from Bow06\\
MHIBow06& Eq.~\ref{st1} &8 &0.62& 485       & N-body     & Modified star formation recipe from Bow06\\
GpcBow06& Eq.~\ref{st1} &4 &0.72& 390       & Monte Carlo & Different background cosmology and \\
        &               &  &    &           &             & modified star formation recipe from Bow06 \\
\hline
\end{tabular}
\end{table*}

Semi-analytical models of galaxy formation invoke simple, physically motivated 
recipes to follow the fate of the baryons in a universe in which structure in the 
dark matter grows hierarchically (White \& Rees 1978; White \& Frenk 1991; 
Kauffmann et~al. 1993; Cole et~al. 1994; for a review of this approach see Baugh 2006). 
The current generation of models include a wide range of phenomena, ranging from 
the heating of the intergalactic medium, which affects the cooling of gas in 
low mass haloes, to the suppression of cooling flows in massive haloes 
due to heating by accretion of matter onto supermassive black holes (e.g. Bower et~al. 2006; 
Croton et~al. 2006; Cattaneo et~al. 2007; Monaco et~al. 2007; Lagos, Cora \& Padilla 2008). 
In this paper, we use the Durham semi-analytical galaxy formation code {\tt GALFORM} 
to make predictions for the amount of cold gas in dark matter haloes of different masses. 
This code was introduced by Cole et~al. (2000) and has been developed in a series of papers 
(Benson et~al. 2003; Baugh et~al. 2005; Bower et~al. 2006; Font et~al. 2008). The code predicts 
a wide range of properties for the galaxy population in the context of a spatially flat 
cold dark matter cosmology with a cosmological constant. 

In this paper we consider four different models run using {\tt GALFORM}. 
Two of these are available from the Millennium 
Archive\footnote{http://galaxy-catalogue.dur.ac.uk:8080/Millennium/}; these are 
the Bower et al. (2006; hereafter Bow06) and Font et al. (2008) models (hereafter Font08). 
The third model is a modified version of the Bow06 model (which we label as MHIBow06), which is discussed 
in more detail below. In this model a small number of parameters have been adjusted from the values 
used in Bow06 in order to produce a better match to the cold gas mass function estimated by  
Zwaan et~al. (2005). The fourth model (denoted by GpcBow06) is set in a different background 
cosmology from the other three, which adopt the cosmology of the Millennium simulation 
(Springel et~al. 2005).
The cosmology of the GpcBow06 model is in better agreement with 
recent measurements of the cosmic microwave background and the large-scale structure of the 
Universe (Sanchez et~al. 2009).\footnote{The cosmological parameters used in the Millennium simulation are 
a matter density $\Omega_{0}=0.25$, a cosmological constant $\Lambda_{0}=0.75$, a Hubble 
constant $H_{0}=73 \,{\rm km s}^{-1}{\rm Mpc}^{-1}$, a primordial scalar spectral index $n_{\rm s}=1$, 
baryon density $\Omega_{\rm b}=0.045$ and fluctuation amplitude $\sigma_{8}=0.9$. 
In the Sanchez et~al. (2009) best fitting model these parameters become $\Omega_{0}=0.26$, 
$\Lambda_{0}=0.74$, $H_{0}= 71.5 \,{\rm km s}^{-1}{\rm Mpc}^{-1}$, $n_{\rm s}=0.96$, 
$\Omega_{\rm b} = 0.044$, and $\sigma_{8} = 0.8$.} 
The Bow06, Font08 and MHIBow06 models use merger histories extracted from the Millennium Simulation. 
The GpcBow06 model uses Monte Carlo generated merger trees as described below. 
When we make predictions for the spatial distribution of galaxies in the GpcBow06 model, we use 
the GigaParsec simulation run at the Institute for Computational Cosmology (GPICC; Baugh et~al. 
in preparation), which uses 10 billion particles to model the hierarchical clustering of mass 
in a simulation cube $1000 h^{-1}\,$Mpc on a side. 
To keep the number of models manageable, we do not consider the Baugh et~al. (2005) model in this paper. 
This model was included in the study by Power et~al. (2009). The star formation recipe used in the 
MHIBow06 model is based on that used in Baugh et~al. (2005).

\begin{figure*}
\includegraphics[width=5.7cm]{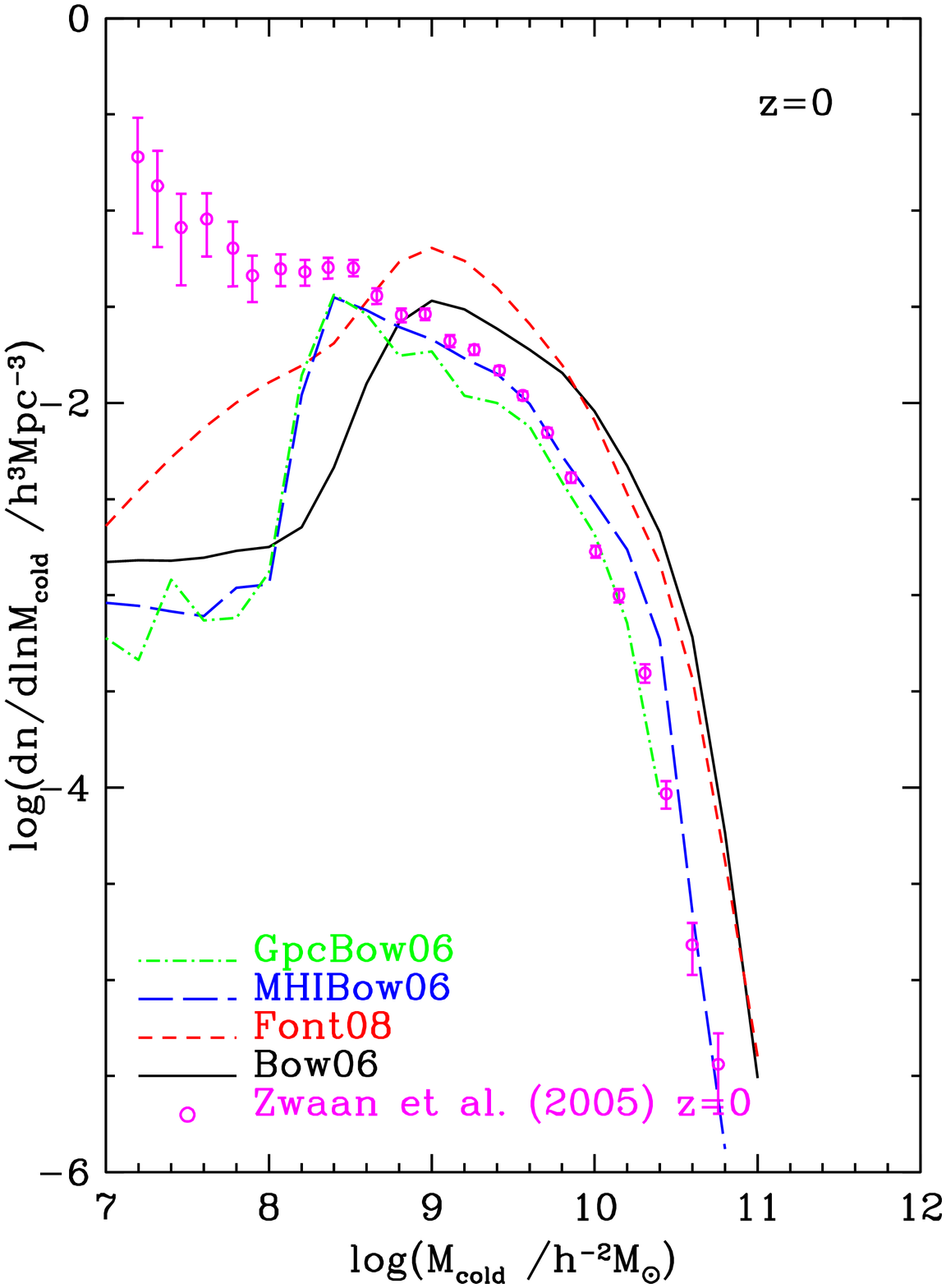}
\includegraphics[width=5.7cm]{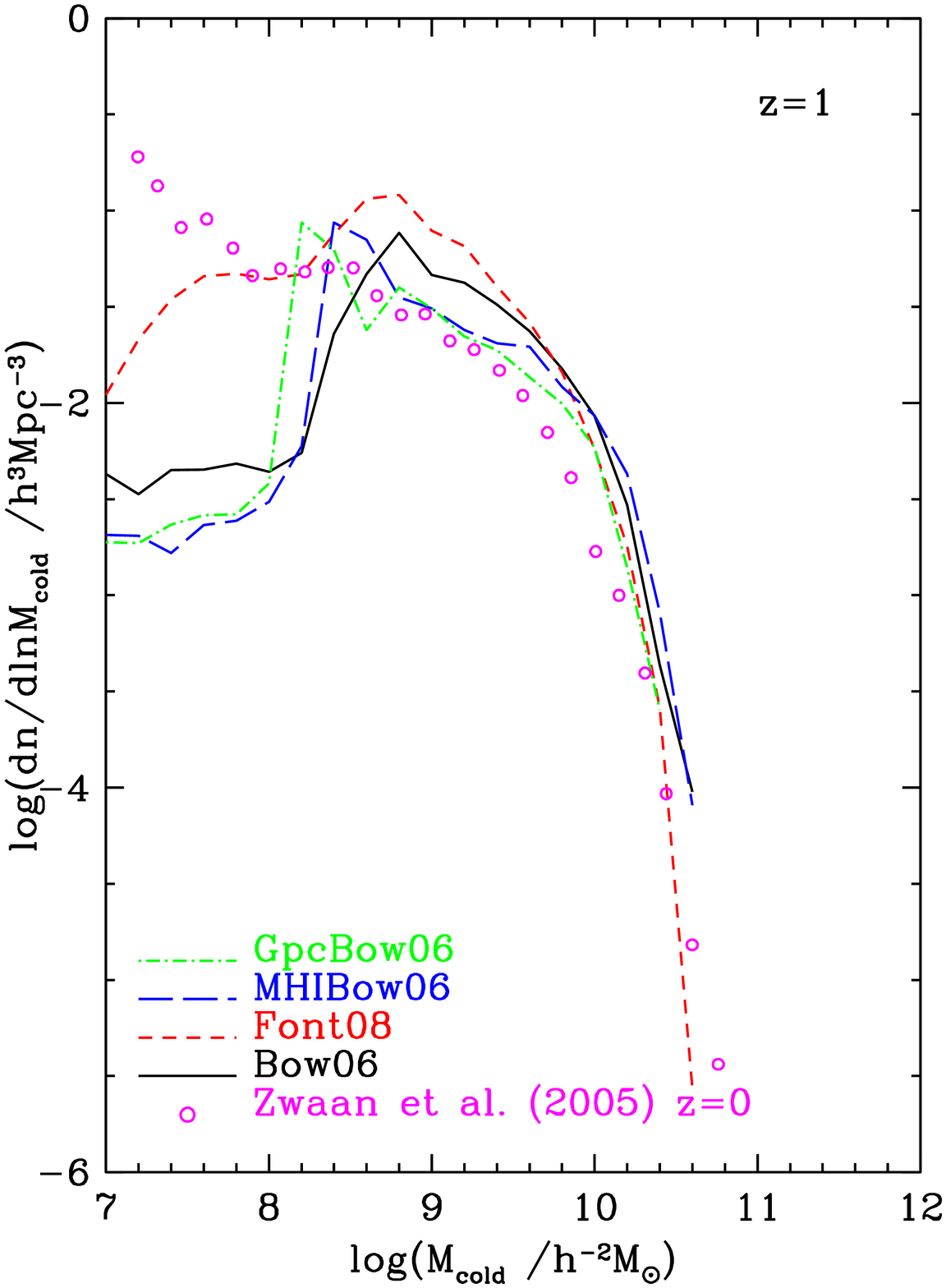}
\includegraphics[width=5.7cm]{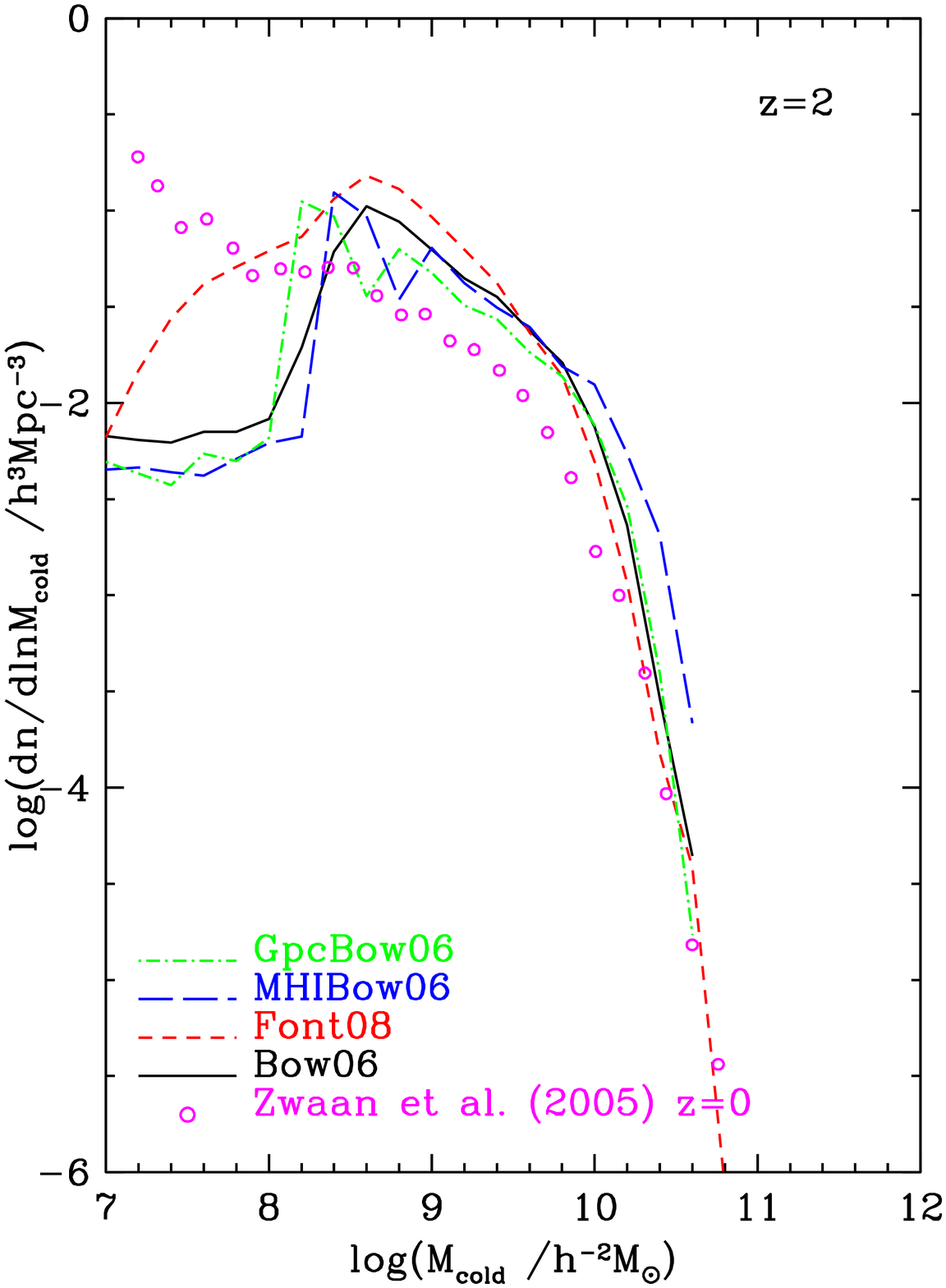}
\caption{
The cold gas mass function predicted in the four models at $z$=0 (left), $z=1$ (middle) and  $z$=2 (right). 
Different colours and line types correspond to different models as indicated by the legend. The points show 
the local ($z=0$) observational estimate of the cold gas mass function inferred from the HI mass function of 
Zwaan et~al. (2005) (see text in Section~\ref{Clustering} for details of the conversion). These data are 
reproduced without error bars in the $z$=1 and $z$=2 panels as a reference from which 
to illustrate the evolution of the mass function.} 
\label{mcold_z0}
\end{figure*}

The MHIBow06 and GpcBow models use the same basic physical ingredients as the Bow06 model. 
The Font08 model is based on Bow06, with a modification to the cooling prescription. 
We discuss these differences in more detail below. We first discuss some of the ingredients 
which are varied between the models, in order to introduce some of the parameter definitions 
used in {\tt GALFORM}. 

All of the models we consider in this paper include the suppression of cooling flows in massive haloes, 
as a result of the energy released following accretion of matter onto a central 
supermassive black hole (Bower et~al. 2006; Malbon et~al. 2007; Fanidakis et~al. 2009).
A halo is assumed to be in quasi-hydrostatic equilibrium if the time required for gas 
to cool at the cooling radius, $t_{\rm{cool}}(r_{\rm{cool}}$), exceeds a multiple 
of the free-fall time at this radius, $t_{\rm{ff}}(r_{\rm{cool}})$: 
\begin{equation}\label{AGN}
t_{\rm{cool}}(r_{\rm{cool}}) > \frac{1}{\alpha_{\rm{cool}}} t_{\rm{ff}}(r_{\rm{cool}}),
\end{equation}
where $\alpha_{\rm{cool}}$ is an adjustable parameter, whose value controls the sharpness 
and position of the break in the optical luminosity function.  
The cooling flow in the halo is then shut down completely if the luminosity released  
by accretion of matter onto the supermassive black hole (SMBH) exceeds the cooling 
luminosity. The energy released by accretion depends on the mass of the SMBH (see, for 
example, Fanidakis et~al. 2009). 

The models also include the ejection of cooled gas into the hot halo due to heating 
by supernovae. The strength of supernovae feedback is defined by the factor $\beta$:
\begin{equation}
\beta=(V_{\rm{hot}}/V_{\rm{disk}})^{\alpha_{\rm{hot}}}.
\label{SNe}
\end{equation}
The rate at which gas is reheated is $\beta$ times the star formation rate.
Here $V_{\rm{disk}}$ is the circular velocity of the disk at its half mass radius, and 
$V_{\rm hot}$ and $\alpha_{\rm hot}$ are parameters. A similar 
equation holds for supernova feedback in the galactic bulge. In {\tt GALFORM}, 
the parameters $V_{\rm hot}$ and $\alpha_{\rm hot}$ are set without reference to 
the number of supernovae. The primary constraints on these parameters are the 
shape of the luminosity function, the slope of the disk rotation speed - luminosity 
relation and the scale size of disks (see Cole et~al. 2000). 

The Bow06 and Font08 models use a star formation timescale, $\tau_{\star}$, which is 
proportional to the galactic dynamical time, $\tau_{\rm{dyn}}$, and is given by :
\begin{equation}\label{st2}
\tau_{\star}=\epsilon_{\star}^{-1}\tau_{\rm{dyn}}(V_{\rm{disk}}/200\,{\rm{km s^{-1}}})^{\alpha_{\star}}, 
\end{equation}
where $\alpha_{\star}$ and $\epsilon_{\star}$ are adjustable parameters ($\alpha_{\star}=-1.5$ in both 
cases). 
The dynamical time is defined as $\tau_{\rm{dyn}} = r_{\rm{disk}}/V_{\rm{disk}}$. 
In contrast, the MHIBow06 and the GpcBow08 models adopt a star formation timescale 
which does not depend on the galactic dynamical time. Instead, in these cases, the timescale is given by :
\begin{equation}\label{st1}
\tau_{\star}=\tau_{\star}^{0}(V_{\rm{disk}}/200\,{\rm{kms^{-1}}})^{\alpha_{\star}},\\
\end{equation}
where $\tau_{\star}^{0}$ and $\alpha_{\star}$ are adjustable parameters (again, in both 
cases, $\alpha_{\star}=-1.5$); this parameterization 
was used in Baugh et~al. (2005). 

The Font08 model includes an improved treatment of the ram-pressure stripping of 
hot-gas atmospheres of satellite galaxies, motivated by the hydrodynamic simulations 
of McCarthy et~al (2008). Also in this model, the yield of metals per solar mass of stars 
formed is increased by a factor of two over the default but rather uncertain value expected 
for a standard solar neighbourhood stellar initial mass 
function. These changes are motivated in part by the desire to improve the predictions 
of the Bow06 model for the colour magnitude relation of central and satellite galaxies 
in groups. The revision to the stellar yield reddens the colour of all galaxies  
in the Font08 model compared with Bow06. The change in the cooling model changes the 
relative abundance of galaxies in the red and blue populations at low luminosities. In the Font08 model, 
there are more faint blue satellite galaxies than in the Bow06 model. These galaxies are 
starved of freshly cooled gas in Bow06 and so had redder stellar populations. The predicted 
colours in the Font08 model are in much better agreement with the observed colour 
magnitude relation measured by Weinmann et~al. (2006). 

The motivation for the MHIBow06 model is clear from Fig.~\ref{COMCOLD}. This plot shows the 
galactic neutral hydrogen mass to optical luminosity ratio and the cold gas mass function 
at the present day. 
Note that when we plot mass function (lower panels of Fig.~\ref{COMCOLD}, 
cold gas masses are plotted in units of $h^{-2}M_{\odot}$ 
rather than $h^{-1}M_{\odot}$, which is the unit used in the simulation. This ensures 
that the observational units (which depend upon the square of the luminosity distance) are matched.
The Bow06 model predicts a gas mass to luminosity ratio with the wrong zeropoint and slope. 
Since this model gives an excellent match to the local optical luminosity function, the discrepancy 
in the gas to luminosity ratio results in a poor match to the cold gas mass function. 
The MHIBow06 model uses the star formation timescale given by Eq.~\ref{st1} and also adopts a 
different value for the AGN feedback free parameter, $\alpha_{\rm{cool}}$ (Eq.~\ref{AGN}; see 
Table~\ref{Parameters}). The right hand panels of Fig.~\ref{COMCOLD} show that the MHIBow06 model 
is in much better agreement with the observed gas to luminosity ratio and cold gas mass function for 
cold gas masses in excess of $\sim 3 \times 10^{8} h^{-2} M_{\odot}$. 
Note that the models predict the mass of cold gas, which includes helium, atomic hydrogen and 
molecular hydrogen. The observed mass function in Fig.~\ref{COMCOLD} is measured in terms of the atomic 
hydrogen (HI) content of galaxies. To convert this into a cold gas mass, we have assumed a 
fixed ratio of molecular to atomic hydrogen and corrected for the mass fraction of Helium 
(see Power et~al. 2009). We shall return to this point in Section 5.  

The GpcBow06 model starts from the Bow06 model, with small adjustments made to the galaxy 
formation parameters to obtain a good match to the optical luminosity function (this is required 
because the cosmological model has changed from that used in Bow06) and also to reproduce the 
observed HI mass function.  The GpcBow06 model uses Monte-Carlo merger trees generated using 
the improved algorithm devised by Parkinson et~al. (2008).

Fig.~\ref{mcold_z0} shows the cold gas mass function predicted by the four models at $z$=0,\,1 and 2. 
The Bow06 and Font08 models overpredict the abundance of galaxies with a given cold gas 
mass at $z=0$ compared with the observational estimate by Zwaan et al. (2005). On the other hand, 
the cold gas mass functions of the MHIBow06 and GpcBow06 models agree well with the local observational 
estimate for masses in excess of $10^{8.5} h^{-2} M_{\odot}$. The discrepancy between the predictions 
and observations at lower masses is not due to the finite resolution of the N-body halo merger 
trees. The turnover can be traced back to the modelling of the photoionisation of 
the intergalactic medium and the impact this has 
on the cooling of gas in low effective circular velocity haloes. In all cases a particularly simple 
approach is taken to model this effect, whereby cooling in low circular velocity haloes ($v_{\rm c} < v_{\rm cut}$) 
is suppressed below the redshift at which the universe is assumed to have been reionised ($z_{\rm cut}$).  
The parameters adopted ($v_{\rm cut} = 50 {\rm \,km\, s}^{-1}$ and $z_{\rm cut} = 6$) may overestimate 
the impact of this effect according to recent simulations by Okamoto, Gao \& Theuns (2008). 
The form of the observed HI mass function at low masses could give interesting contraints on the 
modelling of photoionisation and supernova feedback (Kim et al, in preparation). 
Here we focus on the more massive galaxies which dominate the overall HI content of the 
Universe.

\section{The spatial distribution of cold gas}
We now compare the predictions of the four galaxy formation models for 
the spatial distribution of cold gas with one another and with observations. 
To understand the spatial distribution of cold gas, we first 
look at the halo occupation distribution (HOD; Benson et~al. 2000; Peacock \& Smith 2000; 
Berlind \& Weinberg 2002). This quantifies the number of galaxies above a given 
cold gas mass, as a function of dark matter halo mass (\S~3.1). We present predictions for the 
correlation function of galaxies selected by their cold gas mass in \S~3.2.

\subsection{The halo occupation distribution}
\label{HODs}

\subsubsection{Variation of Cold Gas Mass with Halo Mass}

\begin{figure}
\includegraphics[width=8cm,bb=50 250 580 700]{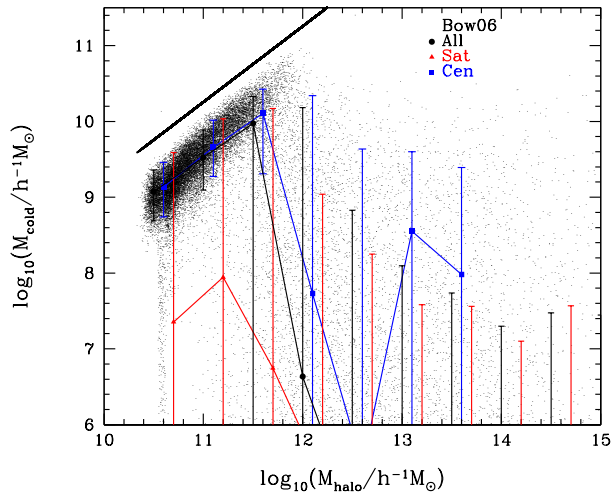}\vspace{0cm}
\caption{\label{MHMC}
The cold gas mass of galaxies in the Bow06 model as a function of the 
mass of their host dark matter halo. The black points show individual 
galaxies. The symbols joined by lines show the median cold gas mass as 
a function of halo mass, for central galaxies (blue), satellite 
galaxies (red) and all galaxies (black). The bars show the 10-90 percentile 
range of the distribution of cold gas masses. All galaxies, including those 
with zero cold gas mass are included when computing the median and percentile 
range. The solid black line shows the cold gas mass a galaxy would 
have if all the available baryons in its halo were in the form of cold gas in 
one object.  
}    
\end{figure}

Before considering the halo occupation distribution directly, 
it is instructive to first look at how the cold gas mass of galaxies 
varies with the mass of their host dark matter halo, which we 
plot in Fig.~\ref{MHMC} for the Bow06 model. The median cold gas mass 
as a function of host halo mass is plotted separately for central 
and satellite galaxies.   
There is a tight correlation between the mass of cold gas of a central galaxy and 
its host halo mass for galaxies in 
haloes less massive than $\sim 3 \times 10^{11} h^{-1}M_{\odot}$. In haloes 
more massive than this, AGN feedback suppresses gas cooling and there is a 
dramatic break in the galaxy cold gas mass - halo mass relation, with an accompanying increase 
in the scatter. The galaxies with the largest mass of cold gas do not 
lie in the most massive dark matter haloes, but reside instead in haloes 
with masses $\sim 10^{12} h^{-1} M_{\odot}$. The predicted cold gas mass - 
halo mass relation is remarkably similar to that inferred 
observationally (Wyithe et~al. 2009a). Another conclusion that is readily 
apparent from Fig.~\ref{MHMC} is that the bulk of the baryons associated 
with a dark matter halo are not in the form of cold gas. The solid line in this 
plot shows the mass a galaxy would have if all of the available baryons in 
the halo were in the form of cold gas in one object, assuming the universal baryon fraction. 
The points are some way 
below this line for two reasons: 1) in most haloes, the bulk of the baryons 
are in the hot phase and 2) there is more than one galaxy in most haloes.

\subsubsection{Cold Gas Halo Occupation Distributions}

\begin{figure*}
\includegraphics[width=16cm]{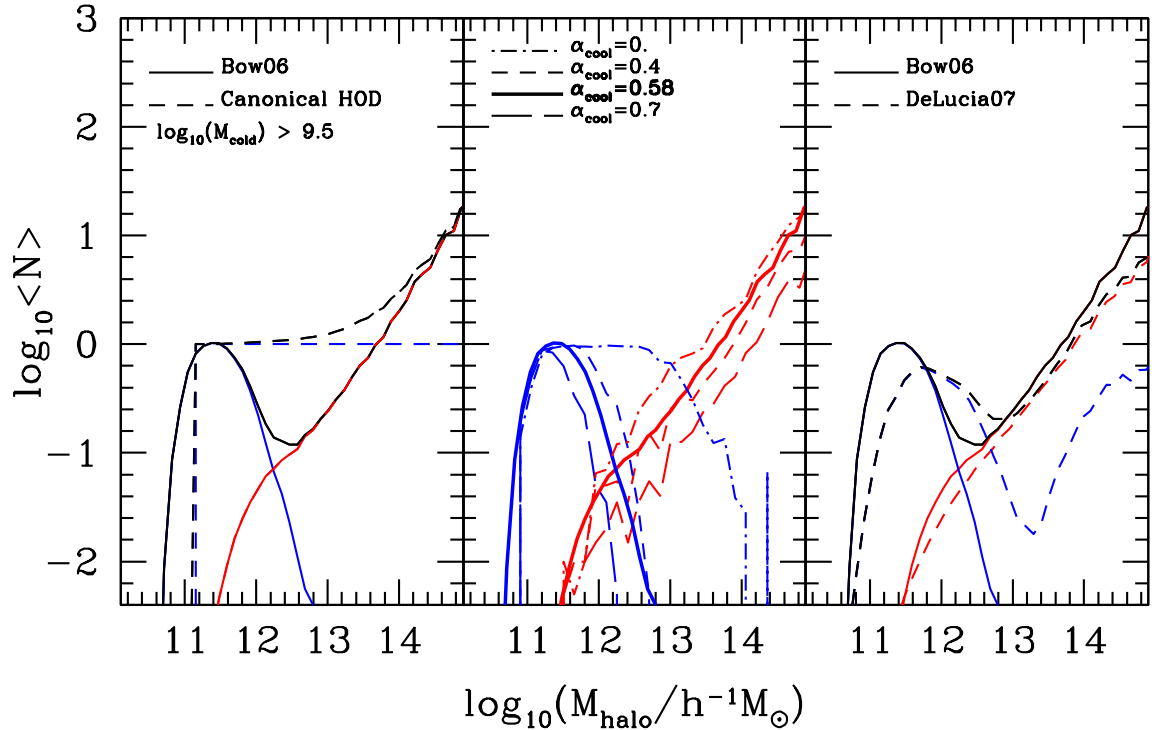}\vspace{-5cm}
\caption{\label{HODAGN}
The predicted halo occupation distribution (HOD) of galaxies with cold gas mass in excess of 
$10^{9.5}{h}^{-2}{\rm{M}}_{\odot}$, chosen to match the sample of galaxies for which 
Wyithe et~al. (2009) estimated the HOD for in HIPASS. The left panel shows the HOD predicted in the 
Bow06 model (solid lines: blue shows the central galaxy HOD, red shows satellites and black 
shows the overall HOD). The dashed blue line shows a step function designed to reproduce the 
number of central galaxies in Bow06. The dashed black line shows this step function combined 
with the model HOD for satellites. The central panel shows the impact on the HOD of changing the 
halo mass above which AGN feedback stops the cooling flow. The fiducial Bow06 model 
corresponds to $\alpha_{\rm cool} = 0.58$. In a model with a larger value of $\alpha_{\rm cool}$, 
the onset of cooling suppression can shift to lower mass haloes; reducing $\alpha_{\rm cool}$ means 
that cooling is only switched off in more massive haloes. The right hand panel compares the HOD predicted 
by Bow06 (solid lines) with that in the model of DeLucia \& Blaizot (2007), for the same cold gas  
mass threshold (dashed lines). The colour coding is the same in each panel.}    
\end{figure*}

We now examine the predictions for the halo occupation distribution (HOD) of galaxy samples 
constructed according to cold gas mass. The HOD gives the mean number of galaxies which 
satisfy a given selection criterion as a function of halo mass, and can be broken down 
into the contribution from the central galaxy in a halo and its satellite galaxies. 
In the case of optically selected galaxy samples, the HOD is commonly described by a step 
function for central galaxies and a power law for satellite galaxies (Peacock \& Smith 2000; 
Seljak 2000; Berlind \& Weinberg 2002; Zheng 2004). Many attempts have been made to interpret  
the clustering of optically selected galaxy samples using the HOD formalism (van den Bosch et~al. 2003; 
Magliocchetti \& Porciani 2003; Zehavi et~al. 2005; Yang et~al. 2005; Tinker et~al. 2007; Wake et~al. 2008; Kim et~al. 2009). 
In contrast, there are few studies of the clustering of galaxies selected on the basis of their 
atomic hydrogen mass using the HOD formalism (Wyithe et~al 2009a, 2009b; Marin et~al. 2009).  

Fig.~\ref{HODAGN} shows the typical form predicted by the models for the HOD of galaxies selected by their cold gas mass. 
The left panel shows the HOD for galaxies in the Bow06 model which have cold gas masses in excess of 
$3 \times 10^{9}{h}^{-2}{\rm{M}}_{\odot}$, chosen to have the same HI mass cut as HIPASS. For this mass threshold, the abundance of central galaxies is 
sharply peaked around a halo mass of $\sim 2 \times 10^{11} h^{-1} M_{\odot}$. The HOD of satellite 
galaxies reaches unity in haloes which are a hundred times more massive. In these haloes, the central 
galaxy has a cold gas mass below the cut-off; there is essentially zero chance of finding a halo which 
contains a central galaxy and a satellite galaxy above this cold gas mass threshold. 
However, this does not imply that it is impossible to find more than one galaxy per halo with 
cold gas masses above the threshold, simply that when this occurs (i.e. once a sufficiently 
massive halo is considered), both galaxies will be satellites. 

For comparison, we also plot in the left hand panel of Fig.~\ref{HODAGN} the traditional form adopted for the HOD of central galaxies 
(i.e. a step function). The minimum halo mass in this case  
is set by the requirement that the step function reproduces the number of central galaxies in the 
Bow06 model. The step function HOD is markedly different to the predicted HOD, which is closer to 
a Gaussian. A similar conclusion about the peaked form of the central galaxy HOD  was postulated by 
Zehavi et~al. (2005) for blue central galaxies. Wyithe et~al. (2009a) model the clustering of 
galaxies in the HIPASS survey by adopting a step function for the central 
galaxy HOD and a truncated power law for satellite galaxies, such that haloes 
above some mass cut contain no satellites. The truncation point lies in the halo 
mass range $10^{14}$-$10^{15} h^{-1} M_{\odot}$, depending on the slope of the satellite HOD. 
As we shall see later on, whilst this truncation is not predicted by any of the models, this has little 
impact on the abundance or clustering of the galaxies.  

In Fig.~\ref{HODAGN}, the HOD of central galaxies in the Bow06 model drops far below unity above a halo mass of 
$\sim$10$^{12}{h}^{-1}{\rm{M}}_{\odot}$. In this model there is very little cold gas in 
haloes more massive than this due to the shut down of the cooling flow by AGN heating. 
To illustrate this, in the middle panel of Fig.~\ref{HODAGN} we vary the halo mass which marks the 
onset of AGN heating by changing the value of the $\alpha_{\rm{cool}}$ parameter (see Eq.~\ref{AGN}). 
Reducing the value of 
$\alpha_{\rm{cool}}$ results in the halo mass in which cooling stops being shifted to higher 
masses. In the absence of AGN heating (i.e. $\alpha_{\rm{cool}}$=0), the central galaxy HOD still 
drops below the unity in the most massive haloes ($M_{halo}$$>$10$^{13}{h}^{-1}{\rm{M}}_{\odot}$) due 
to the longer cooling time of the gas in these haloes. These haloes typically have a lower formation 
redshift and thus a lower gas density and are also hotter; hence they have a longer cooling time. 
Cold gas is depleted by star formation in such massive haloes.

\begin{figure*}
\includegraphics[width=16cm]{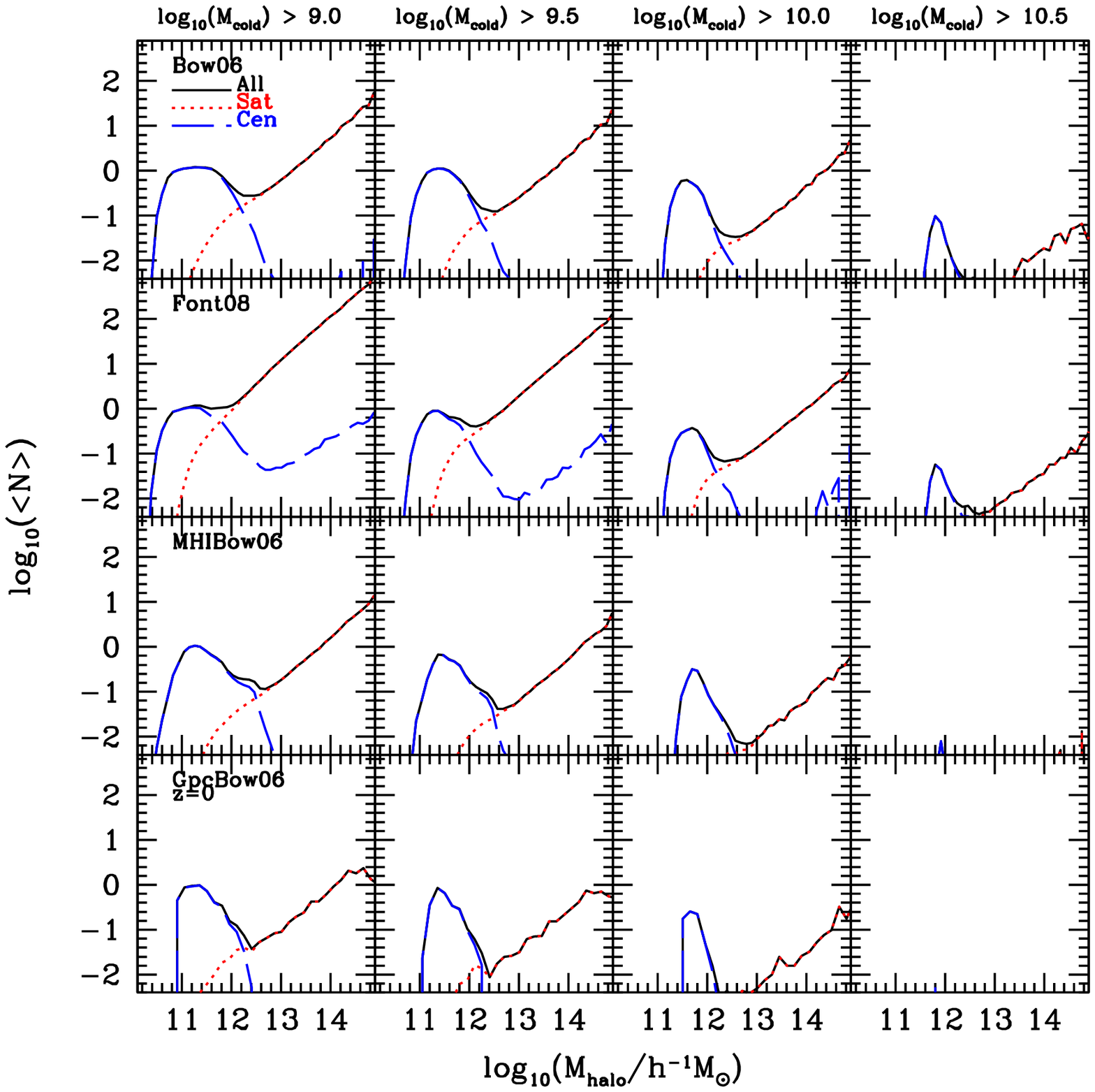}\vspace{-1cm}
\caption{\label{HIHOD63} 
The halo occupation distribution, i.e. the mean number of galaxies passing the selection labelled per halo, 
at $z=0$ for galaxy samples defined by cold gas mass thresholds. 
The blue dashed curves show the contribution from central galaxies, the red dotted curves show satellite 
galaxies and the black solid curves show the overall HOD. Each row corresponds to a different model, 
and each column to a different cold gas mass threshold, as labelled.} 
\end{figure*}

\begin{figure*}
\includegraphics[width=16cm]{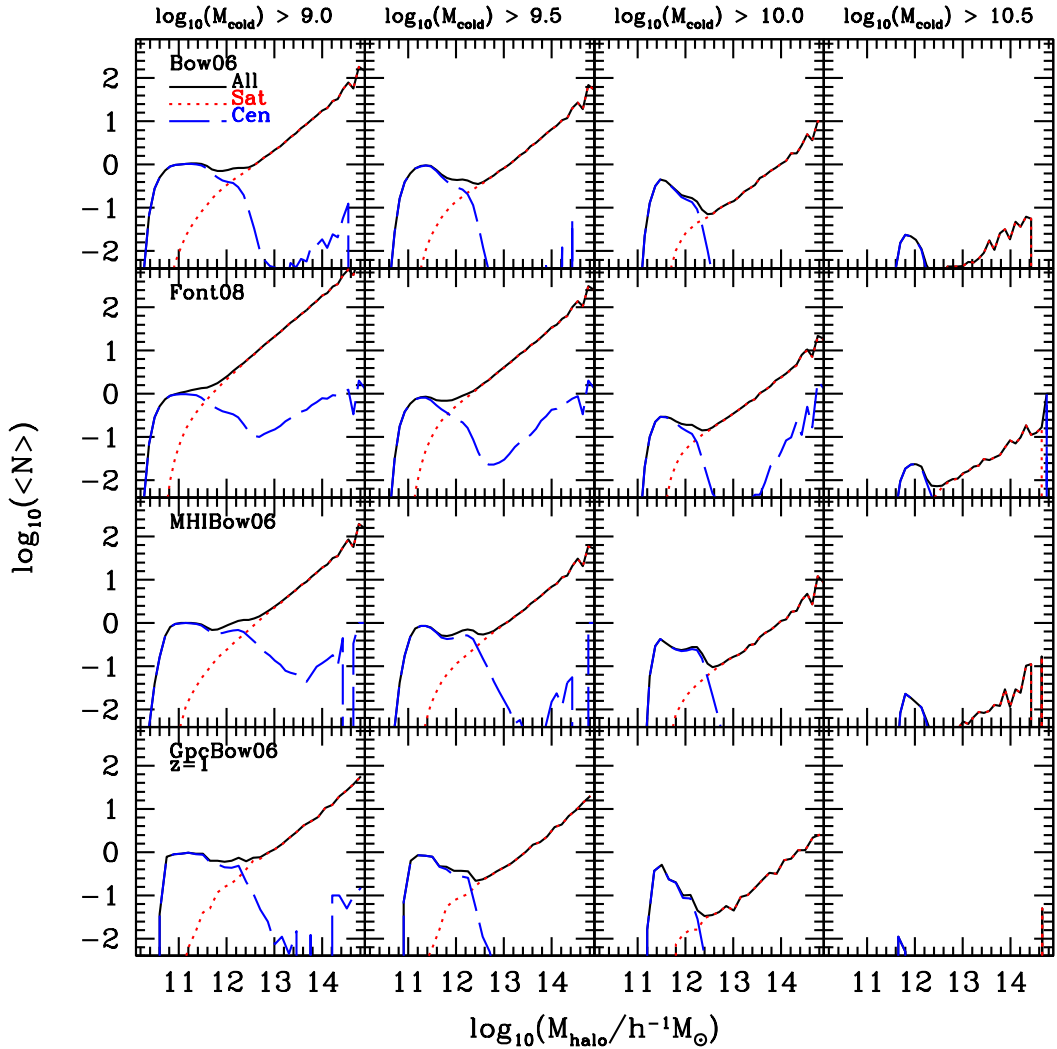}\vspace{-1cm}
\caption{\label{HIHOD41}
The halo occupation distribution at $z=1$ for samples defined by a threshold cold gas mass.  
The blue dashed curves show the contribution from central galaxies, the red dotted curves show 
satellite galaxies and the black solid curves show all galaxies. 
Each row shows a different model as labelled, using the notation set up in Section 2. 
}
\end{figure*}

\begin{figure*}
\includegraphics[width=16cm]{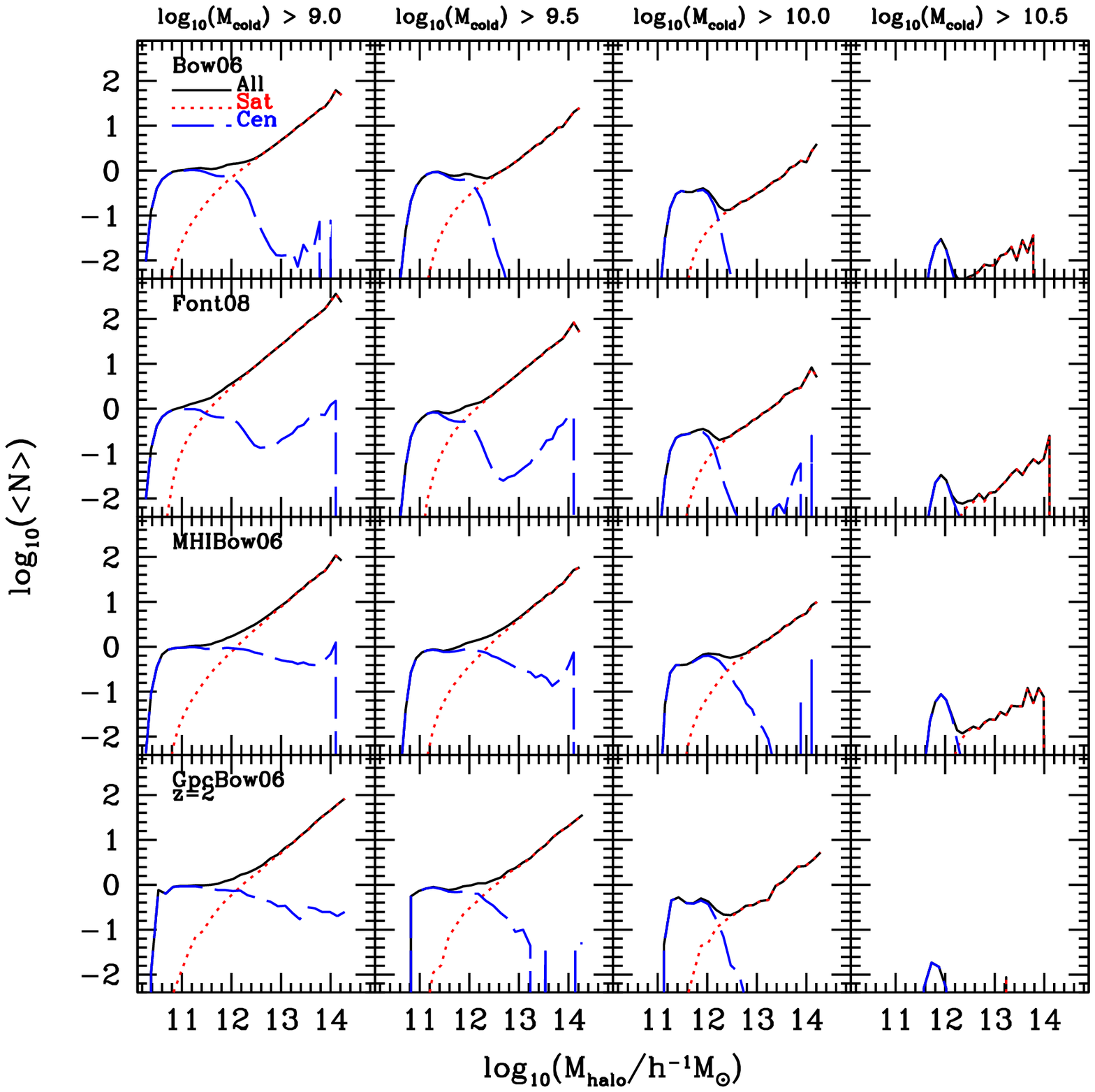}\vspace{-1cm}
\caption{\label{HIHOD32} The halo occupation distribution at $z=2$. As before, the blue dashed 
curves show the contribution from central galaxies, the red dotted curves show satellite galaxies 
and the black solid curves show all galaxies. Each row shows a different model as described in 
Section \ref{model}. Each column corresponds to a different cold gas mass threshold as labelled. } 

\end{figure*}

We shall see later that the peaked HOD for central galaxies is common to all of the {\tt GALFORM} 
models considered, particularly at low redshift. We now examine whether or not this feature is 
peculiar to the way AGN feedback is implemented in {\tt GALFORM} by comparing the Bow06 predictions 
with those of De Lucia $\&$ Blaizot (2007; hereafter the DeLucia07 model). The right hand panel 
of Fig.~\ref{HODAGN} shows that the central galaxy HOD in the DeLucia07 model is somewhat broader than 
that predicted in Bow06, and even increases beyond a halo mass of 
$\sim 2 \times 10^{13} h^{-1} M_{\odot}$. However, as we shall demonstrate further on in this section, 
this upturn has little impact on the predicted clustering. The suppression of gas cooling 
in the DeLucia07 semi-analytical model is smoother than in {\tt GALFORM} 
(see Croton et~al. 2006 for a description of 
the implementation of radio mode feedback). Some gas is permitted to cool 
in haloes with hot gas atmospheres in the DeLucia07 model, with the cooling 
rate modified by accretion onto the central SMBH. 
In {\tt GALFORM}, the cooling flow and heating rate are assumed to balance exactly whenever there is 
a quasi-hydrostatic hot halo and the Eddington luminosity of the black hole exceeds the cooling luminosity. 

Figs.~\ref{HIHOD63}, ~\ref{HIHOD41} and ~\ref{HIHOD32} show the HOD in the four Durham models at 
$z=0, 1$ and $2$. Each column shows the HOD predicted for a different cold gas mass threshold, 
with the mass cut increasing to the right. The rows show the different models introduced in Sec.~\ref{model}. 
For the most massive cold gas mass threshold plotted in Fig.~\ref{HIHOD63}, the mean occupation 
number in the MHIBow06 and GpcBow06 models is less than 1 galaxy per 100 haloes. In the Bow06 model, 
the HOD peaks at a halo mass just under $10^{12} h^{-1} M_{\odot}$, with around 1 in 10 such 
haloes hosting a central galaxy with cold gas mass above the threshold.

The size of the departure from the traditionally assumed step function HOD for central galaxies at $z=0$ 
in Fig.~\ref{HIHOD63} varies in proportion to the ``strength'' of AGN feedback for the Bow06, Font08 
and MHIBow06 models (see Table~\ref{Parameters}). Although the GpcBow06 model has the weakest AGN feedback, 
the deviation from a step function is largest in this case since this model adopts weaker supernovae feedback than 
the other models (as a result of being set in a different cosmology, with a lower density 
fluctuation amplitude). The departure of the central galaxy from a step function form is less pronounced 
at $z=1$ (Fig.~\ref{HIHOD41}). This is because fewer haloes have hot gas haloes and those which do 
host lower mass SMBH (see Fanidakis et~al. 2009 for plots showing how the mass of SMBH is built up 
over time in the models). These trends continue in Fig.~\ref{HIHOD32}, which shows the HOD for the 
{\tt GALFORM} models at $z=2$. The HOD of central galaxies is now better approximated by 
a step function. The HODs become noisy for massive haloes as such objects are extremely rare 
at this redshift. The central galaxy HOD in the Font08 model has a Gaussian form centered on 
halo masses of a few times $10^{11} h^{-1} M_{\odot}$. The HOD displays an upturn for more massive haloes 
which is reminiscent of the HOD in the DeLucia07 model. In Font08, this mass could be 
brought in by merging satellites, which will have a higher cold gas mass than in the other 
Durham models. The central galaxy HOD becomes closer to the canonical step function form 
with increasing redshift. 

Fig.~\ref{HIHOD63} shows that the amplitude of the HOD for satellite galaxies in the 
Bow06 model is higher than in the MHIBow06 and GpcBow06 models. This is due in part to the 
Bow06 model predicting a higher abundance of galaxies by cold gas mass than is observed 
(see Fig.~\ref{mcold_z0}). The Font08 model predicts many more satellite galaxies than the other models 
($\sim$10 times more for the two lowest cold gas mass thresholds). This can be traced back to 
the modified cooling model in Font08, which means that satellites accrete gas that cools 
from their incompletely stripped hot haloes. Also some of the gas which is reheated by supernovae 
in the satellite is allowed to recool onto the satellite rather than being incorporated into 
the main hot halo. The amplitude of the HOD for satellite galaxies at $z=1$ (Fig.~\ref{HIHOD41} 
in the Bow06, MHIBow06, and GpcBow06 models is higher than predicted at $z=0$. Star 
formation depletes the cold gas by $z=0$. The power law slope of the satellite HOD is remarkably 
constant regardless of cold gas mass threshold, redshift or galaxy formation model, with 
$N_{\rm sat} \propto M_{\rm halo}^{0.8}$. The predicted slope is in good agreement with the 
best fitting value determined from clustering in the HIPASS sample, with Wyithe et~al. (2009a) 
reporting a slope of $0.7 \pm 0.4$.

\subsubsection{Comparing HODs for optical and cold gas mass selection}

\begin{figure*}
\includegraphics[width=16cm]{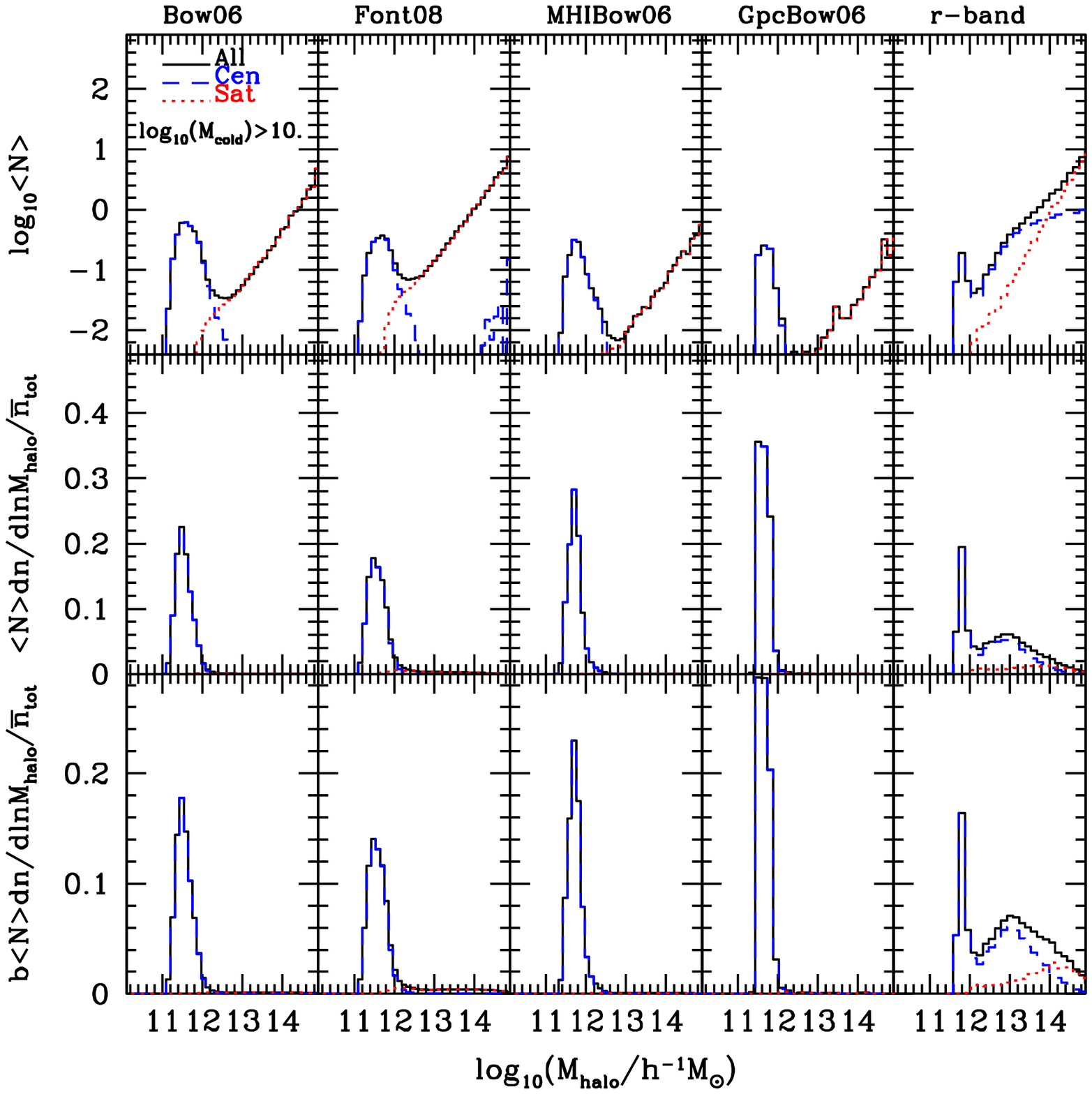}\vspace{-1cm}
\caption{\label{COMHOD5}
The steps relating the number of galaxies per halo to the strength 
of galaxy clustering in the {\tt GALFORM} models. The first row shows 
the HOD as a function of halo mass. The second row shows the HOD 
multiplied by the abundance of dark matter haloes as a function 
of halo mass, $dn/d\ln M_{halo}$, (as computed using the prescription of 
Sheth, Mo \& Tormen 2001) with the $y$-axis plotted on a linear scale. 
$\bar{n}_{\rm tot}$ is the number density of galaxies which satisfy 
the selection criteria (i.e. in cold gas mass or $r$-band luminosity). 
The integral 
of these curves is proportional to the number density of galaxies. The 
bottom row shows the HOD times the halo mass function times the bias factor 
as a function of halo mass . The area under the curves in this case  
gives the effective bias of the galaxy sample. The first four columns show 
the model predictions for galaxies with cold gas mass in excess of 
${\rm{M}}_{\rm{cold}} > 10^{10}{h}^{-2}{\rm{M}}_{\odot}$. 
The fifth column shows an $r$-band selected sample in the GpcBow06 model, 
with the magnitude limit ($M_{r}-5 \log h  < - 21.06$) chosen such that 
the number of galaxies matches that in the cold gas sample in this model. 
As before, the contribution of central galaxies is shown by blue dashed 
lines, satellite galaxies by red dotted lines and all galaxies by black 
solid lines. 
} 
\end{figure*}

We next compare the model predictions for the HOD of an optically selected galaxy sample 
with those of cold gas mass selected samples. Fig.~\ref{COMHOD5} shows the HOD for samples defined by 
a cold gas mass threshold of $10^{10} h^{-2} M_{\odot}$ in the first four columns, with each column 
showing the predictions for a different model. In the right hand column, we plot the HOD for a sample 
in which galaxies are selected on the basis of their $r$-band luminosity in the GpcBow06 model. The 
optical luminosity cut is chosen such that the galaxies brighter than the limit ($M_{r}-5 \log h  < - 21.06$) have the same 
number density as the sample selected by cold gas mass in the GpcBow06 model. 
As we have already remarked, the HODs for the cold gas samples have similar properties, with a peaked 
HOD for central galaxies which declines rapidly with increasing halo mass, and a power law 
HOD for satellites. The HOD for central galaxies in the optical sample shows a local bump for 
haloes masses just below $10^{12} h^{-1} M_{\odot}$, but overall rises gradually, reaching 
unity at a halo mass of $\sim 3 \times 10^{14} h^{-1} M_{\odot}$.  The bump is due to the 
implementation of AGN feedback. The central galaxy HOD drops after the bump as AGN feedback 
``switches on'' in these haloes. Central galaxies hosted by massive haloes are bright in the 
$r$-band, whilst possessing too little gas to be included in the cold gas sample. 

The remaining rows of Fig.~\ref{COMHOD5} show the steps which connect the HOD predictions to 
the effective bias of the galaxy samples, which tells us the clustering amplitude. In the lower 
two rows of this plot we have switched to plotting quantities on a linear scale. 
In the second row of Fig.~\ref{COMHOD5}, the HOD is multiplied by the abundance of the 
host dark matter haloes, giving the contribution to the number density of galaxies as 
a function of halo mass. 
The abundance of the host dark matter haloes is computed using the prescription of 
Sheth, Mo \& Tormen (2001), which gives a good match to simulation results. 
Beyond the break in the mass function, the number of haloes per unit volume drops 
exponentially. This means that satellite galaxies, whose HOD is described by a moderate 
power law, do not contribute significantly to the number of galaxies per unit volume. 
This is true for samples defined either by cold gas mass or $r$-band luminosity. The 
abundance of galaxies is sharply peaked for the cold gas samples. For the optical sample, 
the galaxy number density has a sharp peak just below a halo mass of $10^{12} h^{-1} M_{\odot}$ 
and then shows a broad distribution and an appreciable contribution from  more massive haloes. 
In the bottom row of Fig.~\ref{COMHOD5}, we plot the number density of galaxies multiplied 
by the bias factor as a function of halo mass (as computed using the prescription of Sheth, 
Mo \& Tormen 2001). The square of the bias gives the factor by which the auto-correlation function 
of haloes is boosted on large scales relative to the correlation function of the dark matter. 
The halo bias increases rapidly beyond the break in the mass function, which 
increases the influence of satellite 
galaxies on the effective bias (e.g. Angulo et~al. 2008b). 
Nevertheless, for the cold gas mass samples satellite galaxies still make 
a negligible contribution to the clustering amplitude on large scales, as quantified by the effective bias.
Satellite galaxies make a modest contribution to the effective bias in the $r$-band sample. 
This contribution 
increases if the luminosity cut is made fainter. In summary, the models predict that galaxies with cold gas 
mass in excess of 10$^{10}{h}^{-2}{\rm{M}}_{\odot}$ are predominately central galaxies hosted by dark 
matter haloes of mass 10$^{12}{h}^{-1}{\rm{M}}_{\odot}$. These haloes are less 
massive than the characteristic 
halo mass at $z=0$ in the cosmologies used and so the bias factor of these samples is below unity; 
they are sub-clustered compared to the dark matter. In contrast, the $r$-band sample has an effective bias 
with a significant contribution from more massive haloes which have a larger bias factor. 
The bias factor for the $r$-band selected samples is therefore greater than unity and clustering length 
is larger than it is for cold gas sample (see Fig.~\ref{RERED} later).

\begin{figure*}
\includegraphics[width=8.6cm]{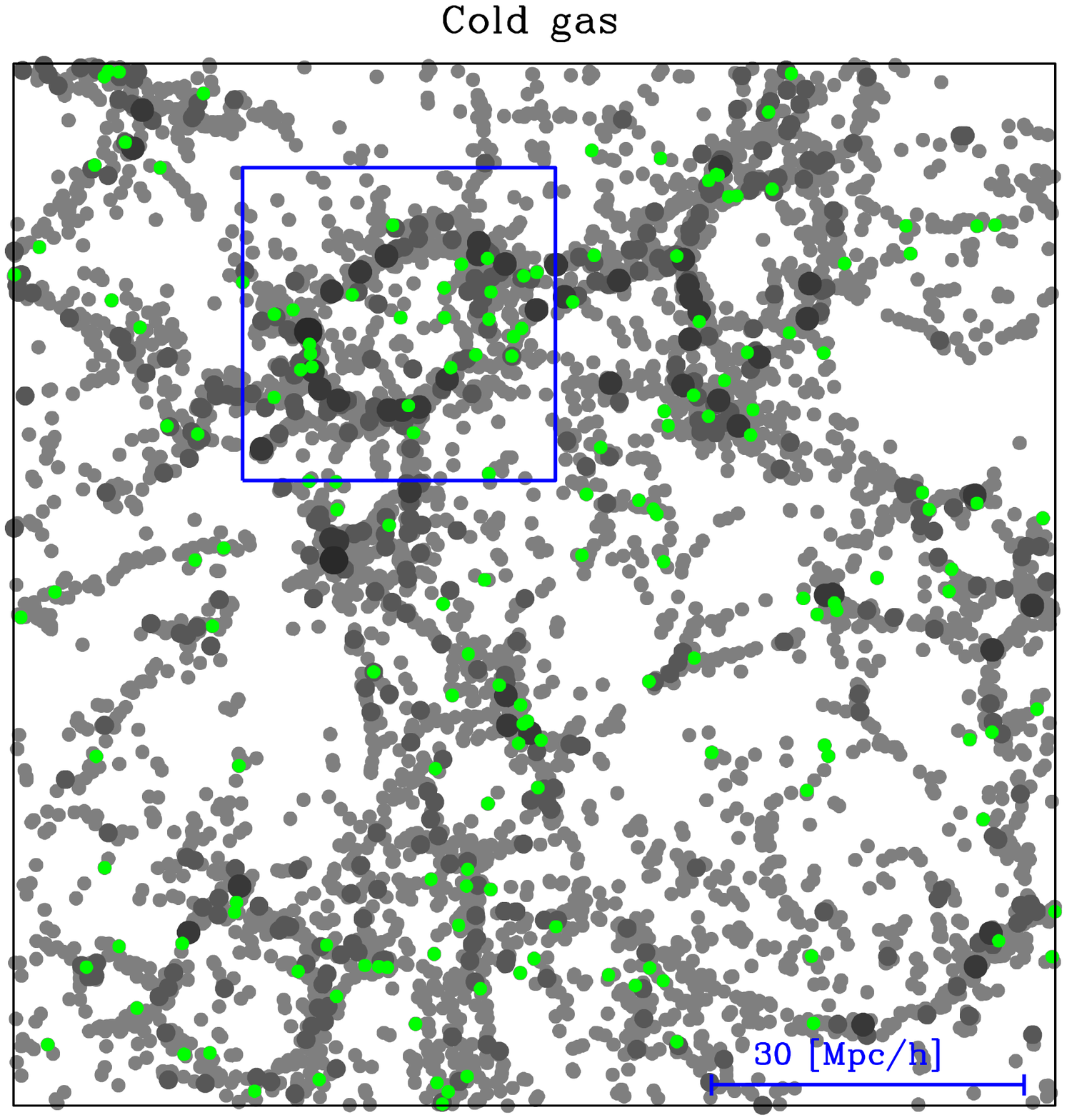}
\includegraphics[width=8.6cm]{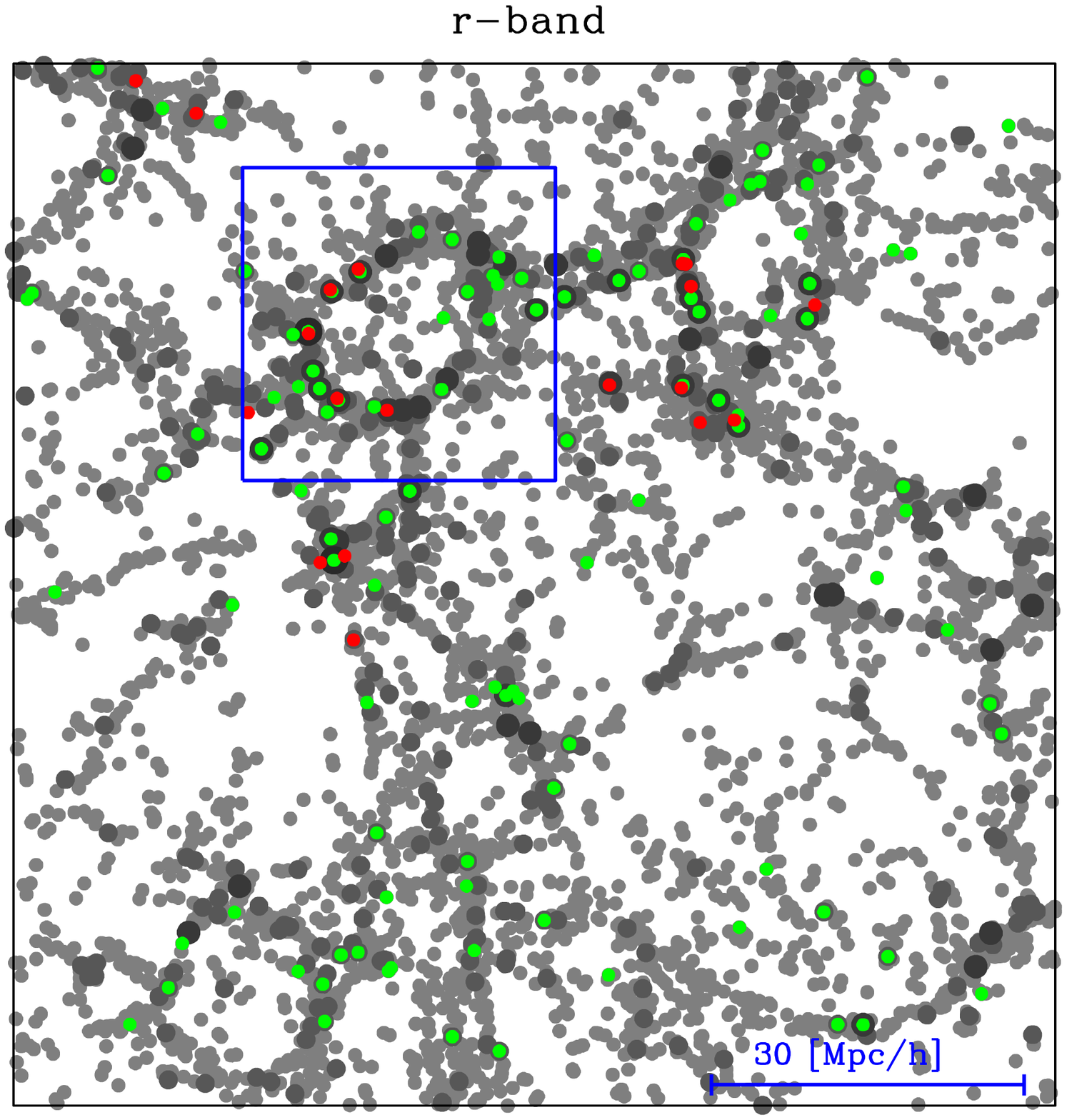}
\includegraphics[width=8.6cm]{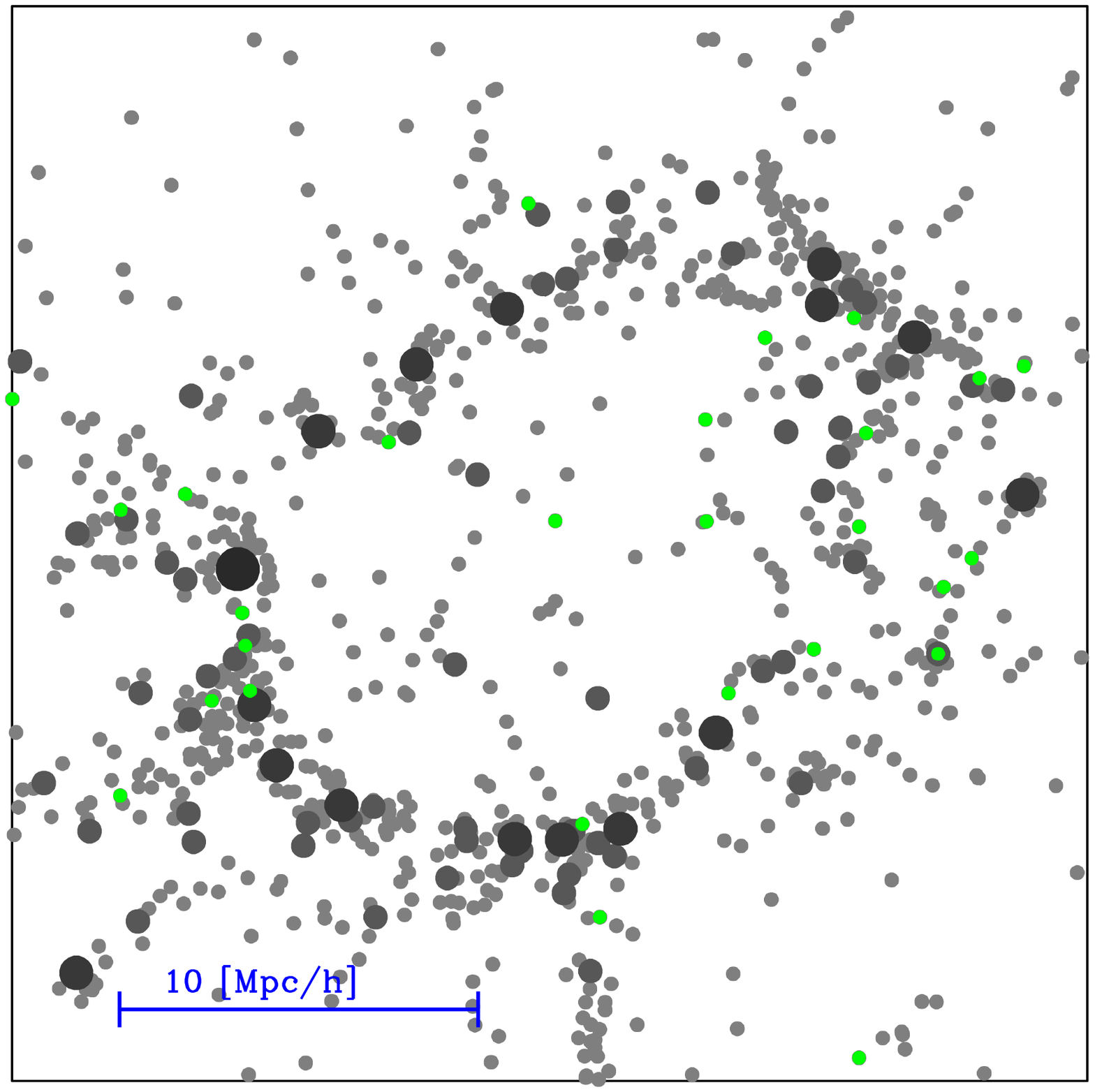}
\includegraphics[width=8.6cm]{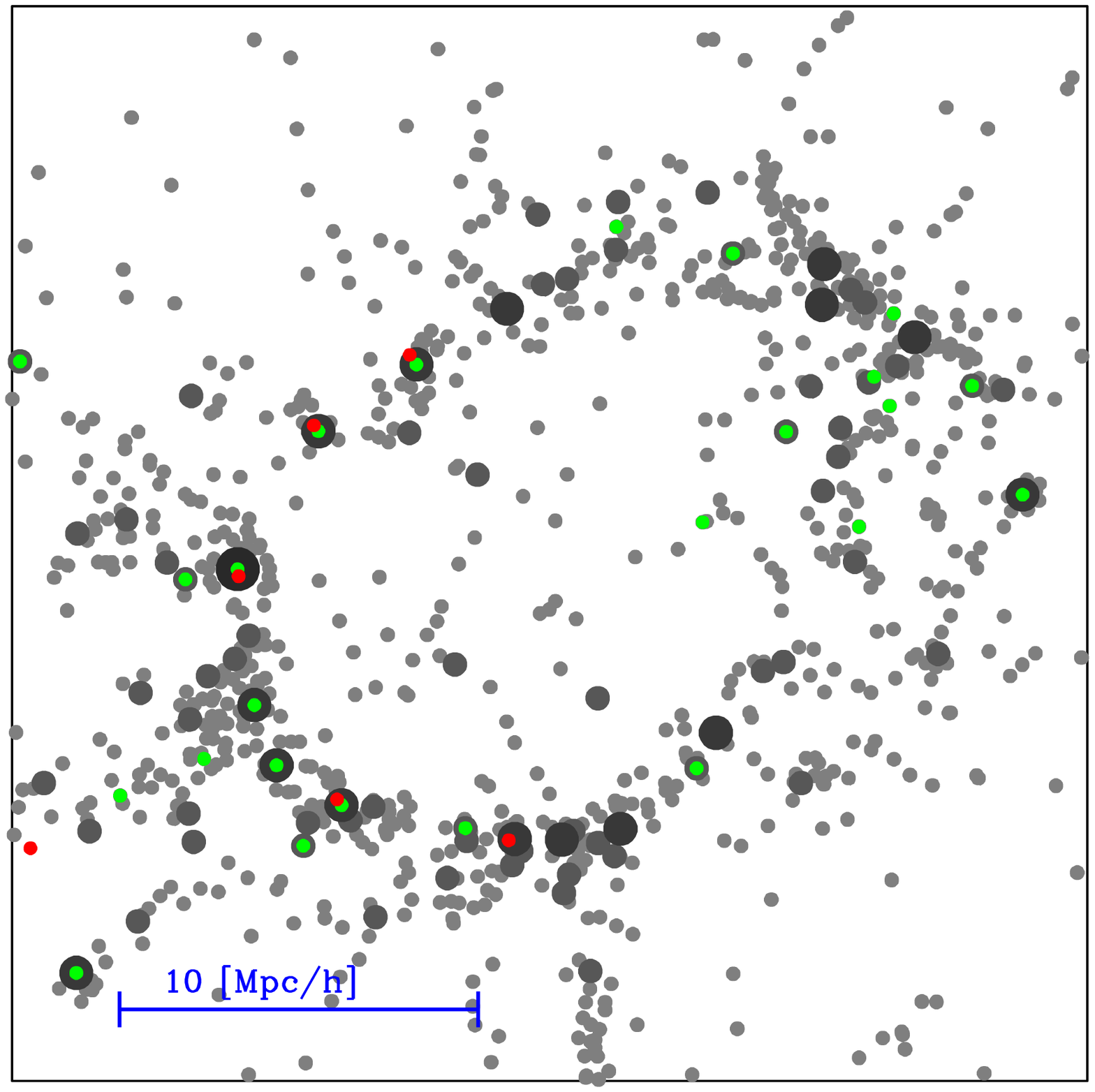}
\caption{\label{DIS}
The spatial distribution of galaxies and dark matter haloes in the GpcBow06 model at $z=0$. 
Dark matter is shown in grey and the size and darkness of the circle used to plot the 
dark matter halo increase with mass. Galaxies selected by cold gas mass 
($M_{\rm cold} > 10^{10} h^{-2} M_{\odot}$) and $r$-band luminosity 
($M_{r} - 5 \log h < -21.06$) are plotted in the left and right hand panels respectively. The top row shows a slice of 100 
${h}^{-1}$Mpc on a side and 10${h}^{-1}$Mpc thick. The bottom row shows a zoom 
into a region of 30${h}^{-1}$Mpc on a side and 10${h}^{-1}$Mpc thick, which 
corresponds to the blue square in the top row. The green circles represent central galaxies 
and the red circles show satellite galaxies.
} 
\end{figure*}

Finally, in Fig.~\ref{DIS} we compare the spatial distribution of $r$-band selected galaxies 
with that of galaxies chosen on the basis of their cold gas mass 
(${\rm{M}}_{\rm{cold}} > 10^{10}{h}^{-2}{\rm{M}}_{\odot}$ in the GpcBow06 model). Again the 
$r$-band magnitude limit ($M_{r} - 5 \log h < -21.06$) is chosen to match the abundance of 
galaxies in the cold gas sample. 
The grey circles represent dark matter haloes. The circle radius and darkness are proportional 
to halo mass. The cold gas selected galaxies follow the filamentary structure and tend to avoid 
high density regions. The difference in the number of satellite galaxies (red circles) is obvious 
between the cold gas and optical samples. The satellites are found in more massive haloes. 
This difference in the spatial distributions provides a visual impression of the differences in 
the HODs plotted in Fig.~\ref{COMHOD5}.
The stronger clustering of the optical samples in principle means that it should be easier 
to measure the power spectrum of galaxy clustering using these tracers. However, the key 
consideration, as we shall see in Section 4, is how the product of the number density of galaxies 
and their power spectrum amplitude changes with redshift. This quantity controls the ``contrast'' 
of the power spectrum signal against the noise which arises from having discrete tracers of the 
density  field.

\subsection{Predictions for the clustering of cold gas}\label{Clustering}

\begin{figure}
\includegraphics[width=8.6cm]{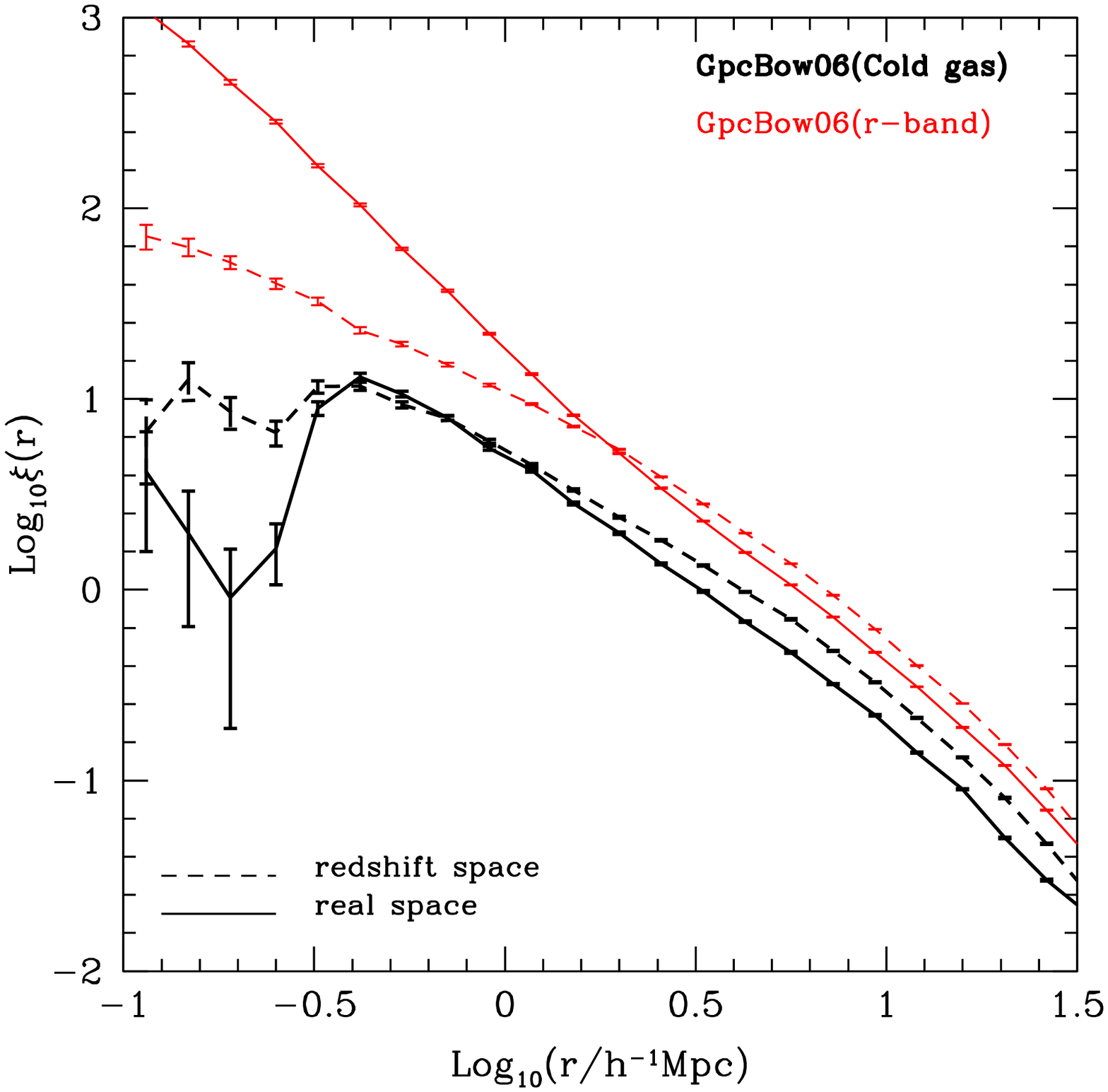}
\caption{\label{RERED}
The real space (solid) and redshift space (dashed) correlation function predicted for 
galaxies in the GpcBow06 model at $z=0$. The black lines show the correlation function 
of galaxies with cold gas mass $M_{\rm{cold}} > 10^{10} h^{-2} M_{\odot}$ and the 
red lines show the clustering of galaxies selected to be brighter than a threshold 
$r$-band luminosity, with the limit chosen to match the abundance of galaxies 
in the cold gas sample. The errorbars show the Poisson error on the pair count 
in each bin of radial separation. 
}
\end{figure}

\begin{figure*}
\includegraphics[width=16cm]{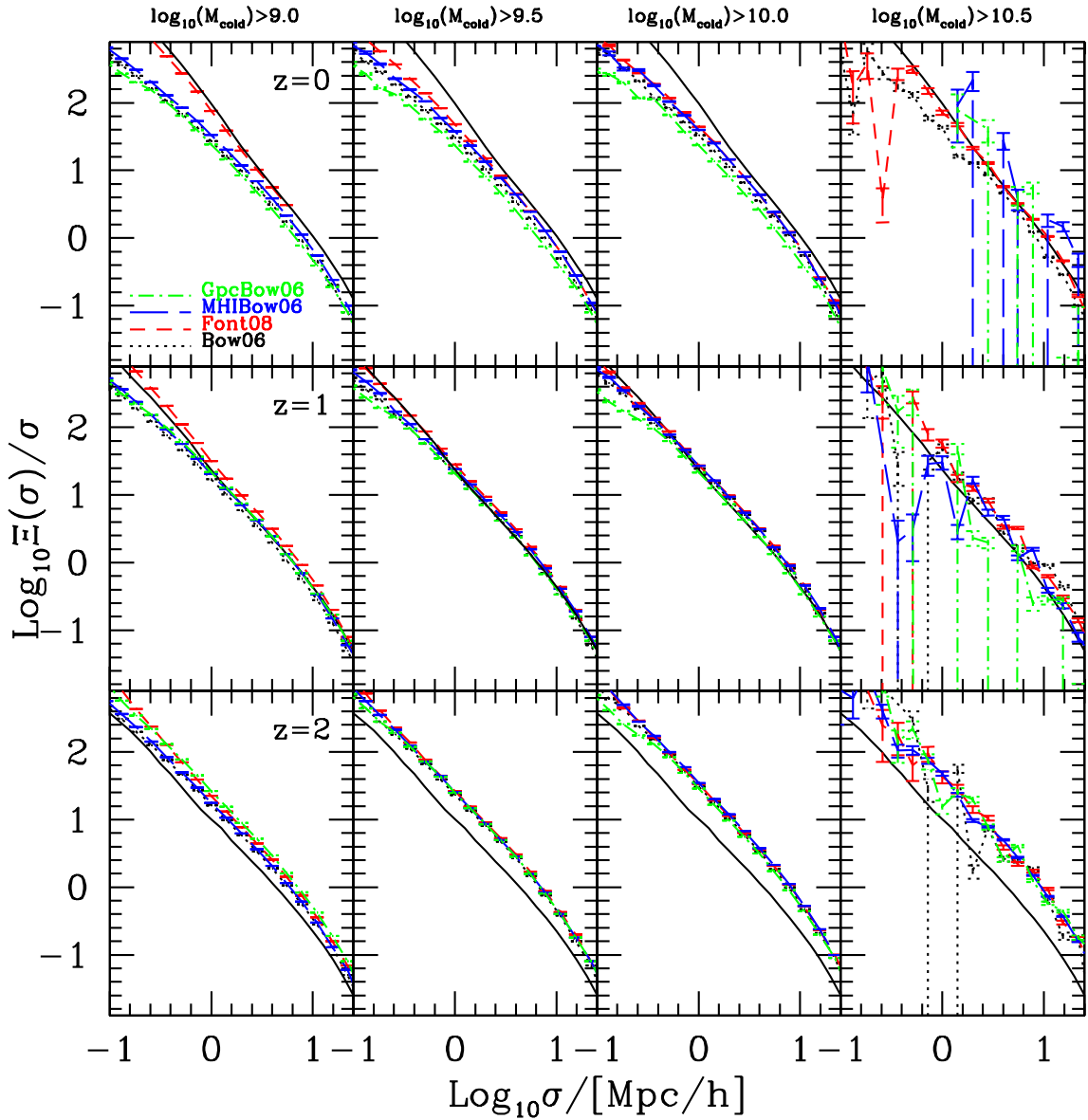}
\caption{\label{HICF63} 
The projected correlation function for cold gas mass selected samples at $z=0$ (top), 
1 (middle) and 2 (bottom). Each column shows the predictions for a different cold gas 
mass threshold, as indicated by the label. The predictions of the models are distinguished 
by different line types and colours, as shown by the key in the upper left panel.
The solid black lines in each panel show the projected correlation function of the dark 
matter measured in the Millennium simulation (note that the GpcBow06 model uses a 
different cosmology and has different dark matter correlation functions).} 
\end{figure*}

\begin{figure}
\includegraphics[width=8.6cm]{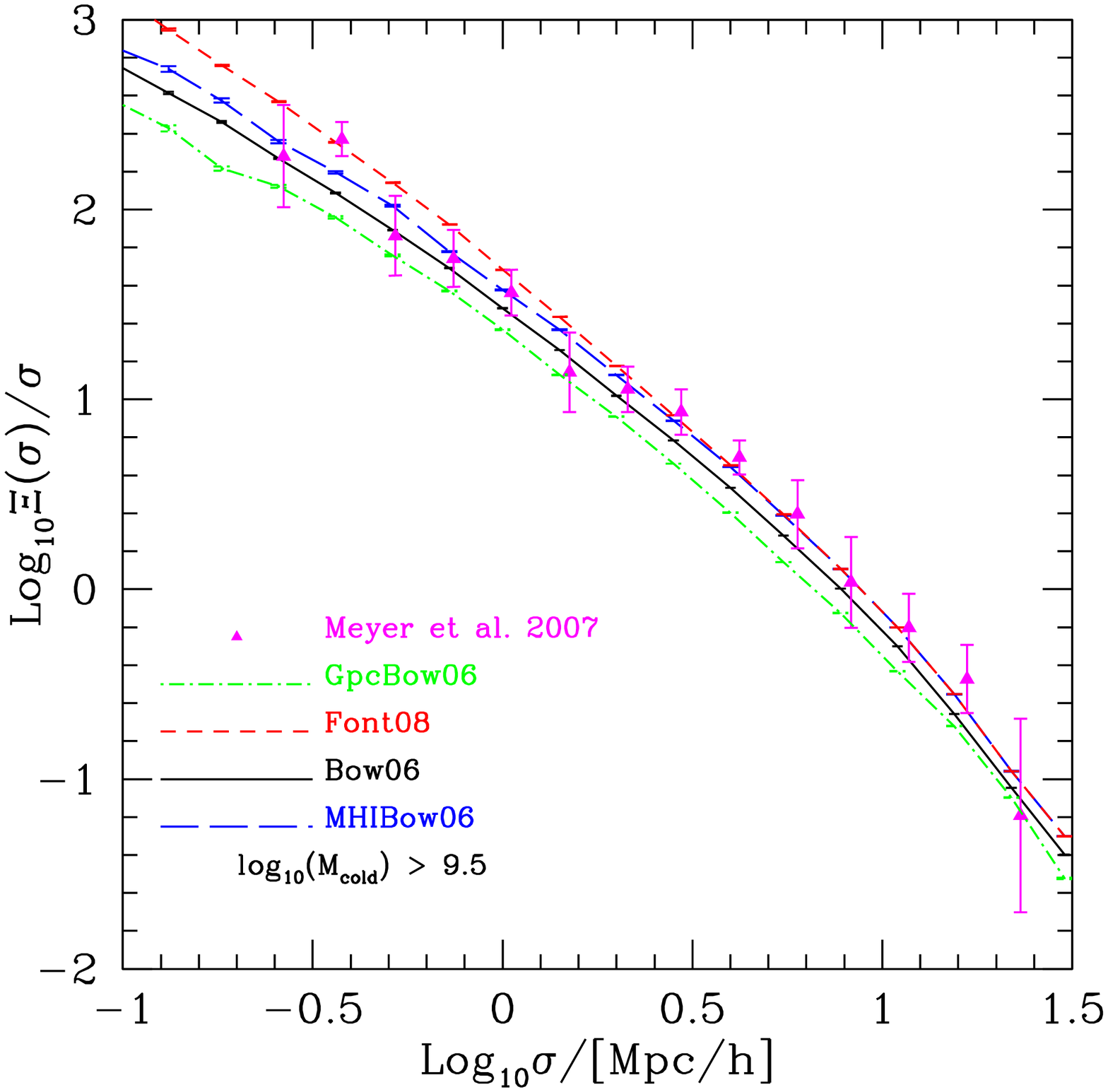}
\caption{\label{MHIDATA}
The projected galaxy correlation function at $z=0$. 
The points with errorbars show an observational estimate made from the 
HIPASS catalogue by Meyer et~al. (2007). 
The lines show the model predictions for galaxies more 
massive than  $M_{\rm{cold}} >  10^{9.5} h^{-2} M_{\odot}$, 
a threshold chosen to match the selection of galaxies in the HIPASS 
sample. The results for different models are shown by 
lines with different colours and line types, as indicated 
by the key. 
} 

\end{figure}

In this section we present the predictions of the galaxy formation models 
introduced in Section~\ref{model} for the two point correlation function. To predict the galaxy distribution of the GpcBow06 model, we generated galaxy samples using the GPICC simulation.

We start in Fig.~\ref{RERED} by comparing the spatial two point autocorrelation 
function of a galaxy sample defined by a threshold cold gas mass 
($M_{\rm{cold}}> 10^{10} h^{-2}M_{\odot}$) in real (solid black line) 
and redshift space (dashed black line). The correlation function is computed 
in redshift space using the distant observer approximation. In this approximation, one of the 
coordinate axes is chosen as the line of sight and the peculiar velocity 
of the galaxy in that direction is added to the real space position, after 
applying a suitable scaling to convert from velocity units to distance units.  
For the largest pair separations plotted, the correlation function in redshift 
space has the same shape as the real space correlation function, but a 
larger amplitude. The magnitude of the shift in amplitude agrees very closely with the 
expectation of Kaiser (1987). This effect is caused by coherent bulk flows 
towards overdense regions. On pair separations between $0.3$ and $1 h^{-1} $Mpc, the 
real and reshift space correlation functions are very similar. They diverge 
on smaller scales, where the predictions are noisy simply because there are 
few galaxies pairs at these separations.   

This behaviour can be contrasted with the clustering in the optically selected 
sample, which is shown by the red lines in Fig.~\ref{RERED}. As with the cold 
gas sample, there is a shift in the clustering amplitude when measured 
in redshift space for pair separations $ r > 3 h^{-1}$Mpc. However, the size 
of the shift is smaller for the optically selected sample, which is consistent 
with the bias of this sample being greater than unity and larger than the bias of the 
cold gas selected sample. The real-space correlation function of the optical sample 
is steep on small scales, reflecting the contribution of satellite galaxies within 
common dark matter haloes. There is a substantial reduction in the clustering amplitude 
in redshift space on these scales in the optical sample, again driven by satellite galaxies. 
This is the so-called ``fingers of God'' redshift space distortion, whereby randomised peculiar 
velocities of the satellites within the gravitational potential of the cluster make the cluster 
appear elongated. 

The real space correlation function cannot be estimated directly from a galaxy redshift 
survey. A related quantity is the projected correlation function which can be estimated from 
the two point correlation function measured in bins of pair separation parallel ($\pi$) and 
perpendicular ($\sigma$) to the line of sight, $\xi(\sigma,\pi)$ (e.g. Norberg~et al. 2001):
\begin{equation}\label{2dcf}
\frac{\Xi(\sigma)}{\sigma} = \frac{2}{\sigma} \int_{0}^{\infty} 
\xi(\sigma,\pi) {\rm d} \pi.
\end{equation} 
In the limit that the integral over the radial pair separation can be taken to infinity, 
this quantity is free from redshift space distortions (see Norberg et~al. 2009 for an 
illustration of the impact of imposing a finite upper limit on the integral).  

Fig.~\ref{HICF63} shows the projected correlation function predicted in the four models 
for a range of cold gas mass samples at $z=0$, 1 and 2. The columns show the results for 
different cold gas mass thresholds, and the rows correspond to different models.
The solid black lines in each panel show the projected correlation function measured for 
the dark matter in the Millennium Simulation (recall that the GpcBow06 model has a different 
cosmology and so should be compared to a consistent dark matter correlation function which 
will be slightly different from that in the Millennium simulation on these scales). 
Overall, the three lowest mass samples at $z=0$ are less clustered than the dark matter. 
The most massive threshold sample we consider at this redshift has a similar clustering 
amplitude to the dark matter. At $z=1$, the bias of the three lowest mass samples is close 
to unity, with the projected clustering of galaxies being very close to that of the dark matter. At $z=2$, 
the cold gas samples are more clustered than the dark matter and correspondingly have 
effective biases greater than unity. This evolution in the bias is due to the adoption 
of a fixed cold gas mass threshold. At high redshift, galaxies with a large cold gas mass 
will tend to be found in more massive haloes. 

Across the different models there is a small spread in clustering amplitude for a given 
cold gas mass sample, with remarkably similar predictions made for the projected 
correlation function. Fig.~\ref{HICF63} shows that the differences start to appear at 
$z=1$ and become larger by $z=0$. The model which shows the largest difference from 
the others is Font08. On small scales in the two lowest mass threshold samples, this model 
has an appreciably higher amplitude projected correlation function than the other models. 
This feature can be traced back to the HODs plotted in Fig.~\ref{HIHOD63}. Due to the 
revised cooling model used in Font08, there are more satellite galaxies in the low mass 
samples in this model, which boosts the one halo term in the correlation function.   

Finally, we compare the predicted correlation functions with an observational estimate from 
Meyer et~al. (2007), which was made using the HI Parkes All-Sky Survey (HIPASS) Catalogue 
(HICAT; Meyer et~al. 2004). In order to make this comparison, we need to convert the cold 
gas mass output by the models into an atomic hydrogen mass. We assume that 76\% by mass of 
the cold gas is hydrogen. Here we adopt a fixed ratio of molecular (H$_{2}$) 
to atomic (HI) hydrogen of H$_{2}/$HI=0.4 (see Power et~al. 2009 for a discussion). 
The HI mass, $M_{\rm HI}$,  is therefore obtained from the cold gas mass $M_{\rm cold}$ by applying 
the conversion: 
\begin{equation}
M_{\rm{HI}}=0.76 M_{\rm{cold}}/(1+0.4).
\end{equation} 
With this relation, the sample analyzed by Meyer et~al. is equivalent to a cold gas mass threshold 
of $M_{\rm{cold}} >  10^{9.5} h^{-2} M_{\odot}$. The comparison between the model predictions 
and the observational estimate is presented in Fig.~\ref{MHIDATA}. 
The correlation function predicted by the MHIBow06 model agrees remarkably well with the 
observational estimate. The GpcBow06 and Bow06 models predict too low a clustering amplitude. 
The Font08 model gives a reasonable match on intermediate and large scales, but somewhat 
overpredicts the clustering amplitude on small scales, hinting that this model has too many 
gas rich satellites in massive haloes.

\section{Measuring dark energy with future HI redshift surveys}\label{BAO}

\begin{figure}
\includegraphics[width=8.6cm]{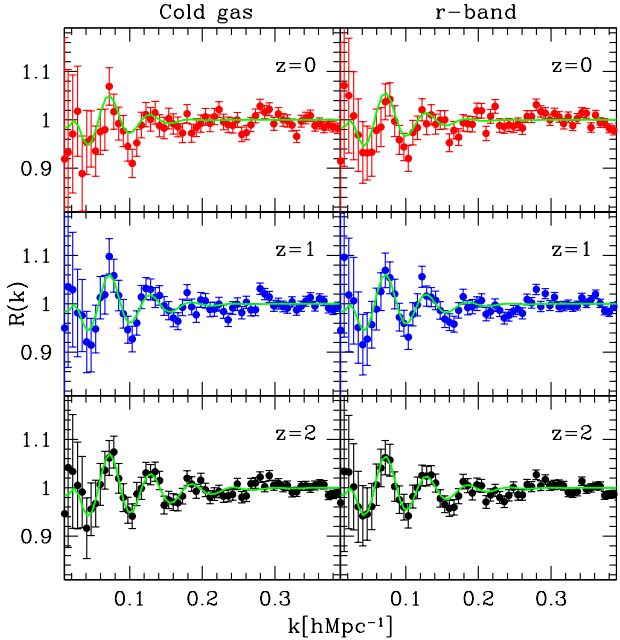}
\caption{\label{BAOGPC}
The baryonic acoustic oscillations in the galaxy power spectrum. To display the BAO 
more clearly, we have divided the predicted spectra by smooth fits, as described in 
the text. The points show the power spectra predicted by the GpcBow06 model at $z=2$ (bottom), 
$z=1$ (middle) and $z=0$ (top). The left hand column shows the power spectra measured 
for galaxies with cold gas mass (${\rm{M}}_{\rm{cold}} > 10^{10}{h}^{-1}{\rm{M}}_{\odot}$). 
The right hand columns shows the BAO in a sample selected in the $r$-band with the same 
number density of galaxies as the cold gas sample at that redshift. 
The smooth green line shows the linear theory power spectrum, divided by a smooth reference 
power spectrum, after filtering or ``de-wiggling'' to damp the higher harmonics (see text).  
The errors plotted on the power spectrum depend on the number density of galaxies and 
the simulation volume (see eq.~3 in Angulo et~al. 2008a). 
} 
\end{figure}

\begin{figure}
\includegraphics[width=8.6cm]{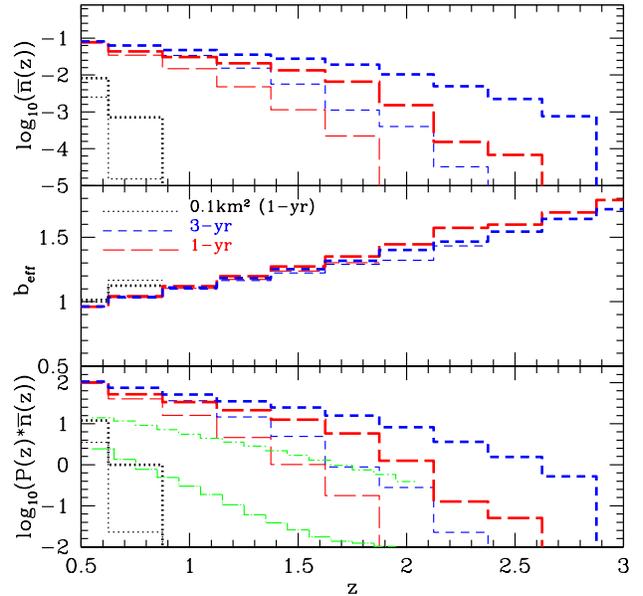}
\caption{\label{BIASSKA}
The quantities needed to compute the effective volume of a redshift survey, as 
predicted in the GpcBow06 model. The upper panel shows the number density of HI 
galaxies for different SKA configurations, with the collecting area and survey 
duration given in the legend. The red and blue histograms show predictions for the 
full SKA collecting area. The field of view is assumed to be 200 square degrees 
in all cases, with the full survey covering one hemisphere. 
Heavy lines show the model predictions for a fixed $H_{2}/$HI conversion, with thin 
lines of the same colour and style showing the predictions for a variable $H_{2}/$HI 
ratio. The middle panel shows the effective bias as a function of redshift for the 
corresponding cases. The lower panel shows the product of the galaxy number density 
and galaxy power spectrum. The value of $P_{\rm gal}(k)$ at $k=0.2 h\, $Mpc$^{-1}$ is plotted. 
The green curves show the predictions for a spectroscopic survey down to $H=22$ (assuming a 
33\% redshift success rate; green dot-long-dashed line) and a slitless survey of H-$\alpha$ down 
to a flux limit of $5\times 10^{-16} $erg \,s$^{-1}$ \,cm$^{-2}$, again with a 33\% redshift 
measurement rate (green dot-short-dashed line); both these results are taken from Orsi et~al. (2009).
}
\end{figure}

\begin{figure}
\includegraphics[width=8.6cm]{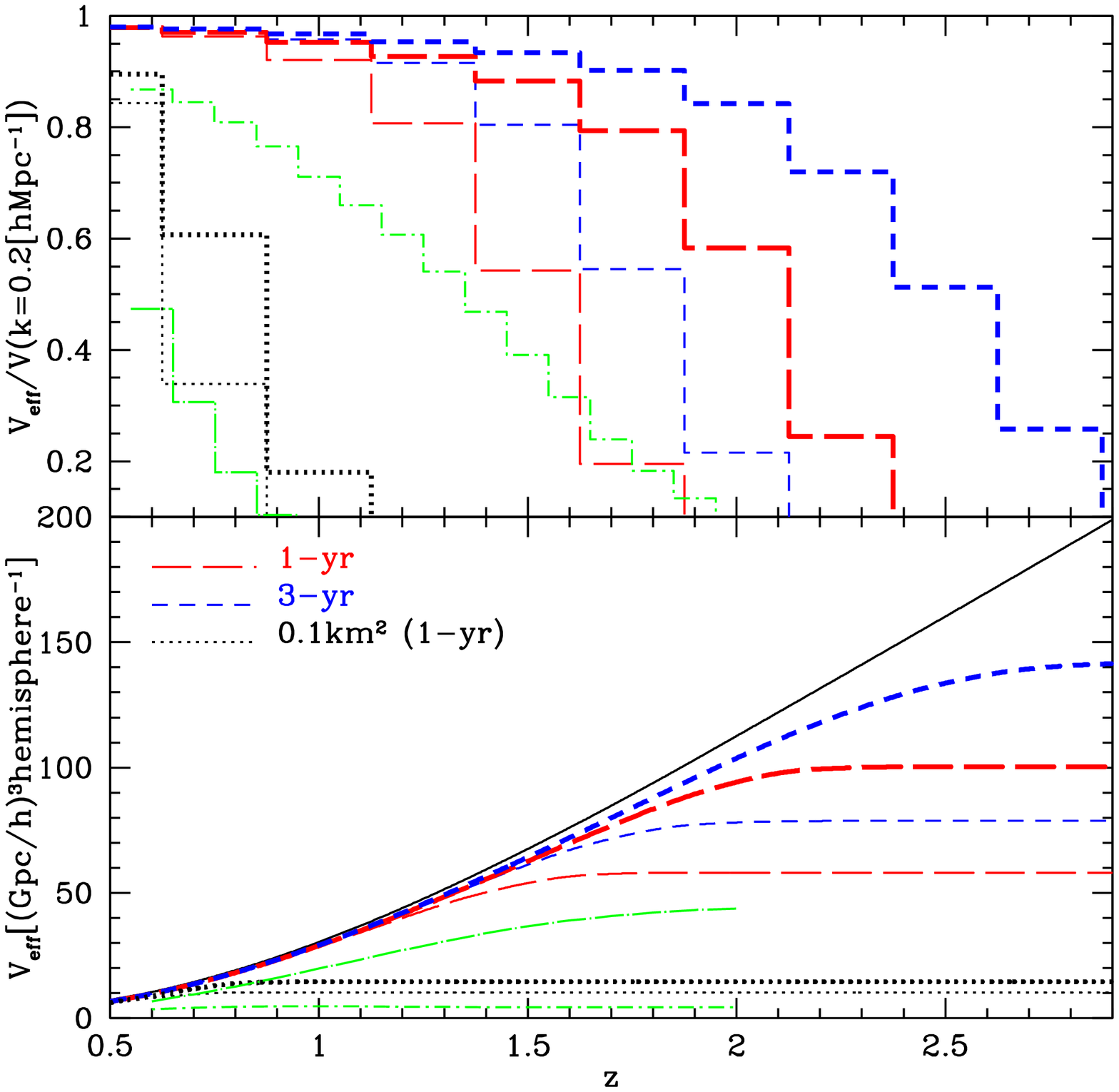}
\caption{\label{VOLUMESKA}
The effective volume per steradian of HI selected samples predicted in the 
GpcBow06 model. The upper panel shows the differential effective volume divided 
by the geometrical volume for narrow bins in redshift. The lower panel shows the 
cumulative volume. The results for different SKA configurations are shown by 
different line styles and colours as indicated by the key. The green curves show 
the predictions for a spectroscopic survey down to $H=22$ (33\% redshift success rate; 
green dot-long-dashed line) and a slitless survey down to an H-$\alpha$ flux limit of $5\times 10^{-16}$ erg s$^{-1}$ cm$^{-2}$ 
(green dot-short-dashed line), as computed by Orsi et~al. (2009). The black solid line in the lower 
panel shows the available geometrical volume per steradian. The optical and HI surveys 
are assumed to cover approximately the same solid angle, one hemisphere. 
} 
\end{figure}

In this section we show how redshift surveys of HI selected galaxies can be used to detect  
baryonic acoustic oscillations (BAO) in the galaxy power spectrum, and we assess the 
relative performance of HI and optical surveys in measuring the large scale structure 
of the Universe.

The BAO signal measured in a sample defined by cold gas mass is shown in Fig.~\ref{BAOGPC}. 
We use the galaxy distribution in the GPICC simulation generated using the GpcBow06 model. 
To show the BAO more clearly, we have divided the measured spectrum by a reference power 
spectrum which contains no wiggles. For the linear theory prediction, which is shown by the 
curves in Fig.~\ref{BAOGPC}, the reference is based on the ``no wiggle'' parametrization 
of the power spectrum given by Eisenstein \& Hu (1998). The no wiggle prediction includes 
the impact of a non-zero baryon component on the width of the turn-over in the matter 
power spectrum. 
The ratio of the linear theory power spectrum, $P^{\rm L}(k)$, to 
the no wiggle prediction, $P^{\rm L}_{\rm nw}$, is ``de-wiggled'' by 
damping the oscillations to represent the impact of nonlinear 
growth and redshift-space distortions 
(e.g. Eisenstein et~al. 2005; Sanchez, Baugh \& Angulo 2008): 
\begin{equation}
\label{pbest}
R_{\rm lin }(k)= 
\left(
\frac{P^{\rm L} }{ P^{\rm L}_{\rm nw} }-1
\right)
\times \exp \left(- \frac{ k^{2} }{  2 k_{\rm nl }^{2}} \right) + 1 ,
\end{equation}
where $k_{\rm nl}$ is the damping scale and is treated as 
a free parameter. 

The overall shape of the power spectra measured from the simulation is 
different from the linear theory prediction due to the nonlinear growth 
of fluctuations and redshift-space distortions (see Angulo et~al. 2008a for 
a step by step illustration of these effects). 
We model this change in shape by multiplying the no-wiggle version of 
the linear theory spectrum by a third order polynomial:
\begin{equation}\label{Distor}
P_{\rm{g}}(k)=(1+ Ak+Bk^{2}+Ck^{3})P^{\rm L}_{\rm{nw}}(k).
\end{equation}
The free parameters $A,B$ and $ C$ are chosen to give the best match to the 
overall shape of the measured power spectrum. All points up to $k = 0.4 h {\rm Mpc}^{-1}$ 
were included in the fit and given equal weight. This approach is more 
straightforward and robust than using a spline fit to a coarsely binned 
measured spectrum, which is sensitive to the number of $k$-bins used. 

We show in Fig.~\ref{BAOGPC} the BAO signal in the GpcBow06 model at $z=0,1$ and 2 
for galaxies selected by their cold gas mass (${\rm{M}}_{\rm{cold}} > 10^{10}{h}^{-1}{\rm{M}}_{\odot}$; 
left column), as an illustration of how a cold gas mass selected sample traces this 
large-scale structure feature. In the right-hand columm of Fig.~\ref{BAOGPC}, we compare 
this with the BAO signal expected for an $r$-band selected sample of galaxies that have 
same number density at each redshift as the cold gas sample.  
The reference power spectrum is defined as described above, using the 
third order polynomial fit to the measured spectrum in each case. 
Fig.~\ref{BAOGPC} shows that we should be able to measure the BAO feature just as 
well using a sample selected by cold gas mass as with an optically selected sample.

Many ongoing and proposed redshift surveys have the goal of determining the 
nature of dark energy by measuring the BAO signal in the galaxy power spectrum. 
A powerful way to compare the expected performance of different surveys for measuring 
large-scale structure is to estimate their effective volumes (see, for example, 
Orsi et~al. 2009). This is essentially an indicator of the ``useful'' survey 
volume which determines the size of the errorbar on the measured power spectrum. 
The effective volume is defined as (Feldman, Kaiser $\&$ Peacock 1994)
\begin{equation}
V_{\rm{eff}}(k,z)=\int^{z_{\rm{max}}}_{z_{\rm{min}}}
\left[{\bar{n}(z)P_{\rm{g}}(k,z) \over 1+\bar{n}(z)P_{\rm{g}}(k,z)}\right]^{2}{ {\rm d}V \over {\rm d}z}
{\rm d}z \,
\end{equation}\label{Veff}
where all quantities are expressed in comoving coordinates and $ {\rm d}V/ {\rm d}z $ 
is the differential comoving volume.
To calculate the effective volume, we therefore need to know the number density of 
galaxies ($\bar{n} (z)$) down to a given survey flux limit and the effective bias ($b(z)$), both as functions  
of redshift. In this calculation, we obtain the galaxy power spectrum using the linear relation 
between galaxy bias and the dark matter power spectrum: $P_{\rm g}(k,z) = P_{\rm {dm}}(k, z=0)b^{2}(z)D^{2}(z)$, 
where $P_{\rm{g}}$ is the galaxy power spectrum, $P_{\rm{dm}}(k,z=0)$ is the linear 
theory dark matter power spectrum at $z=0$, $b(z)$ is the effective bias, and 
$D(z)$ is the growth factor of the dark matter. 

To make predictions for the effective volume of the SKA, we need to convert the cold gas mass 
predicted by the models into an HI line flux, which we do following the prescription set out 
in Power et~al. (2009). A key step is the assumption about the fraction of neutral hydrogen 
which is in molecular form as opposed to atomic hydrogen. Power et~al. (2009) adopted two 
prescriptions: a fixed fraction of 40\% as used by Baugh et~al. (2005) and a variable fraction 
as used by Obreschkow \& Rawlings (2009), based on work by Blitz \& Rosolowsky (2006), 
in which this ratio can vary from galaxy to galaxy. 
We shall refer to these two scenarios as the fixed and variable 
$H_{2}/$HI ratio cases. Power et~al. (2009) showed that the high redshift tail of the count distribution 
in the variable $H_{2}/$HI case is substantially suppressed compared with the fixed $H_{2}/$HI ratio case. 

We calculate the effective volume for possible SKA survey configurations using the GpcBow06 model. 
The model gives the number density of galaxies at different redshifts brighter than 
the specified flux limit and the effective bias of these galaxies. The flux limit 
for a given collecting area and integration time is computed as described in Power et~al. (2009). 
As explained above, we follow the procedure given by Power et~al. (2009) to predict the flux of 
galaxies at 21cm and hence the number which can be detected with a given SKA set-up. 
Fig.~\ref{BIASSKA} shows the predictions for the quantities required to calculate the 
effective volume for various SKA configurations. We calculate the effective volume for 
three different survey integration times and collecting areas assuming a 200 deg$^{2}$ field 
of view. The top panel of Fig.~\ref{BIASSKA} shows the number density of sources, $\bar{n}(z)$, 
as a function of redshift. The galaxy number density for a 3 year survey with a 1\,km$^{2}$ 
collecting area is nearly constant up to $z=2$; for an integration time of just 1 year with the 
same collecting area, the number density of galaxies detected declines rapidly beyond $z \sim 2$. 
If the collecting area is much smaller, 0.1km$^{2}$, a 1-yr survey for SKA probes $z<1$.  

The middle panel of Fig.~\ref{BIASSKA} shows that the bias in the three cases described above changes 
by a much more modest amount than the number density of galaxies does, increasing by 50\% with 
redshift over the range plotted. The increase in effective bias cannot therefore compensate for the 
dramatic drop in the abundance of galaxies in the high redshift tails of the distributions. The 
effective volume of a survey configuration no longer increases with redshift once the product of the galaxy number density 
and the galaxy power spectrum drops below unity. In this regime, the power spectrum signal is swamped 
by shot noise ($P_{\rm shot} = 1/\bar{n}$) arising from the use of discrete galaxies to trace 
the continuous density field, and does not contribute to the statistical power of the survey. 
The product $\bar{n}P$ 
is plotted in the lower panel of Fig.~\ref{BIASSKA}. The different survey configurations track 
the geometrical volume available until the redshift at which $\bar{n}P < 1$. This is clear from 
the lower panel of Fig.~\ref{VOLUMESKA}, in which the effective volume curves flatten once this 
redshift is reached. The thick lines show the effective volume expected for a fixed $H_{2}$/HI ratio. 
The thin lines show that the effective volume sampled drops by a factor of two when a variable  
$H_{2}$/HI ratio is adopted. 

We include in Fig.~\ref{BIASSKA} two predictions for redshift surveys conducted in the 
near-infrared taken from Orsi et~al. (2009), who followed the same procedure we have 
set out above, but for different galaxy selection criteria. 
The predictions for a redshift survey to $H=22$ with a 33\% redshift sampling rate 
(green dot-long-dashed line) and for a slitless survey to an H-$\alpha$ flux limit 
of $5\times 10^{-16}$ erg s$^{-1}$ cm$^{-2}$, again with a 33\% redshift measurement rate 
(green dot-short-dashed line), are plotted for comparison. Leaving aside cost considerations 
and technical feasibility, this comparison shows that a 1-year SKA survey is comparable 
to the one third sampling $H=22$ survey, and samples around 5 times the volume of the 
H-$\alpha$ survey. 

\newcommand{\kperp}{k_{\perp}}
\newcommand{\kpara}{k_{\parallel}}
\newcommand{\krperp}{k_{{\rm ref}_{\perp}}}
\newcommand{\krpara}{k_{{\rm ref}_{\parallel}}}

The effective volume gives a broad brush view of the potential performance of a survey. 
In order to get a more quantitative impression, we need to make a forecast of the error 
on the parameter of interest, which in our case is the dark energy equation of state 
parameter, $w$. This will allow us to assess if the volume sampled by the survey is at 
a redshift which is useful for constraining the value of $w$. The conclusions will depend to 
some extent on the dark energy model adopted. The fiducial model we use is a flat 
cold dark matter universe with a cosmological constant. The cosmological constant has 
little impact on cosmological distances above $z \approx 1.5$-2. Hence, a difference in  
effective volume between survey configurations at these redshifts is likely to have little 
impact on how well $w$ can be measured. This behaviour could change if we adopted a different 
dark energy model, such as one with appreciable amounts of dark energy at early epochs 
(see, for example, the plots of Hubble parameter and luminosity distance in Jennings et~al. 2010).  

To make the forecast of the error on $w$ for a particular survey configuration, 
we use a Fisher matrix approach, closely following the calculation in Seo \& Eisenstein (2003). 
Our goal is to compare the different survey configurations, so we use a number of approximations to 
simplify the calculation. In particular, we work in the flat sky approximation, ignore the impact of 
redshift space distortions on the appearance of the BAO and ignore any evolution of the power spectrum 
over bins of redshift of width 0.1.  
Under these assumptions, the Fisher matrix (for arbitrary parameters) obtained from 
the power spectrum is given by (Tegmark, Taylor \& Heavens 1997; Seo \& Eisenstein 2003),
\begin{eqnarray}
  F_{ij} & = \sum_{i=1}^{N_z}\int_{k_{\rm min}}^{k_{\rm max}} 
  \frac{\partial\ln R(k,z_i)}{\partial p_i} \frac{\partial\ln R(k,z_i)}{\partial p_j}\, \\  \nonumber 
  & \times V_{\rm eff}(k,z_i)\, \frac{4\pi k^2 dk}{2(2\pi)^3},
\end{eqnarray}
where $R$ is the measured power spectrum divided by a smooth reference, as given by Eq.~\ref{pbest} and 
the effective volume, $V_{\rm eff}(k,z)$ is given by Eq.~\ref{Veff}. 
The integration is over the wavenumber interval $k_{\rm min} = 0.02 h {\rm Mpc}^{-1}$ to $k_{\rm max} = 0.2 h$~Mpc$^{-1}$. 
To isolate the cosmological constraints which come from the BAO scale, we ignore any information stored in the 
amplitude of the power spectrum and assume the power spectrum is sensitive to $w$ only through 
the observed angular and radial distance scales.  
The explicit dependence, as given in Seo \& Eisenstein (2003) is,
\begin{eqnarray}
  P_{\rm obs}\left(\krperp, \krpara,z\right) 
  & = \frac{D_A^2(z)_{\rm ref}\,H(z)}{D_A^2{z}\,H(z)_{\rm ref}} \\ \nonumber 
  & P_{\rm true}\left(\sqrt{\kperp^2 + \kpara^2},z\right),
\end{eqnarray}
where $\krperp\equiv \kperp D_A(z)/D_A(z)_{\rm ref}$ and $\krpara\equiv \kpara H(z)_{\rm ref}/H(z)$ relate 
the wavenumbers inferred via an assumed cosmological model and the true physical scales in the power spectrum.  

Our calculation is idealised since we hold the values of the other cosmological parameters fixed. 
Angulo et~al. (2008) showed that making such an assumption can have an impact on the size of the 
uncertainty inferred on $w$ given an error on the BAO distance scale. Here we are interested 
in the {\it relative} error on $w$ between different survey configurations, which we assume are robust to 
whether or not we vary other parameters. We find that the error on $w$, $\Delta w$, is fairly insensitive 
to the assumption about the ratio of $H_{2}/$HI. Even though the tails of the redshift distribution of 
HI emitters are significantly different in these cases, and the effective survey volumes differ, this has 
little impact on $\Delta w$. Also, there is little difference in the accuracy achievable with 1 year 
and 3 year surveys. Finally, we compare the performance of HI redshift surveys with that of surveys in the near 
infrared. The $H=22$ survey is predicted to yield an equivalent $\Delta w$ to the 1 year HI SKA survey. 
The H-alpha survey is expected to give a $\Delta w $ that is twice as large as an $H=22$ survey. 

\section{Summary and conclusions}\label{Summary}

The cold gas content of galaxies and its variation with halo mass lie at the 
core of the galaxy formation process. The amount of cold gas in a galaxy is 
set by the balance between a number of competing processes. The cold gas supply 
comes from the cooling of gas from the hot halo and the accretion of cold gas 
following mergers with other galaxies. Star formation and supernova feedback act 
as sinks of cold gas. Semi-analytical simulations model all of these processes 
in the context of structure 
formation in the dark matter and so are ideally suited to make predictions for 
the distribution of cold gas in haloes of different mass. Since the models can 
make a wide range of predictions, their parameters are set by the requirement that a 
variety of observed galaxy properties be reproduced, not just the local HI data. 
The model predictions can be tested by measurements of the clustering of HI-selected 
galaxy samples, and are necessary to plan surveys to measure the large-scale 
structure of the Universe with the next generation of radio telescopes. 

In this paper we have compared the predictions for the distribution of cold gas in dark matter haloes 
of four versions of the Durham semi-analytical galaxy formation model, {\tt GALFORM}. 
The Bower et~al. (2006) and Font et~al. (2008) 
models are publicly available from the Millennium Archive. These models overpredict 
the local abundance of galaxies as a function of their cold gas mass. This excess was 
straightforward to fix, with the primary adjustment made to the model star formation 
timescale. This modified model, based on Bower et~al. (2006) was still able to reproduce 
the quality of match to the optical luminosity functions enjoyed by Bower et~al. We also 
considered a galaxy formation model set in a different cosmology, to take advantage of 
a N-body simulation with a large enough box size to accurately model baryonic acoustic 
oscillations. This model also adopted a modified star formation timescale to better 
match the local HI mass function. 

The model predictions have several features in common. In agreement with 
observations, satellite galaxies are relatively unimportant in samples 
selected by their cold gas mass. This is true even in the Font et~al. 
(2008) model in which satellites retain some of their hot haloes, depending 
on their orbit within the main halo, and can continue to accrete cooling gas. 
Samples constructed according to a cold gas mass threshold are dominated by 
central galaxies in haloes around $10^{11} h^{-1} M_{\odot}$. The halo 
occuption distribution of central galaxies is peaked in halo mass, rather 
than being a step function, as is the case for optical samples. As the cold 
gas mass cut is increased, the width of the central galaxy HOD increases and the 
amplitude drops. The peaked nature of the HOD of central galaxies is due to 
suppression of gas cooling in masses haloes following heating by AGN. We found the 
same general form for the HOD in a model by de Lucia \& Blaizot 2007, 
in which the implementation of AGN/radio mode feedback is different from that in 
{\tt GALFORM}. 

The relative importance of central and satellite galaxies has an impact on the 
form of the predicted correlation function. The 
correlation function of a galaxy sample selected by cold gas mass is remarkably 
similar on small scales in real and redshift space. For pair separations in 
excess of a few Mpc, the redshift space correlation function has a higher amplitude 
than in real space, as expected given the effective bias of the sample (Kaiser 1987). 
In contrast, for an optically selected sample with the same number density of galaxies, 
the correlation steepens in real space for $r < 1 h^{-1}$Mpc and is damped in redshift 
space on these scales, due to the greater influence of satellite galaxies in massive 
haloes. On larger scales there is a more modest boost in the clustering amplitude in 
redshift space, due to the larger effective bias of the optical sample. 
The clustering predictions are in reasonable agreement with the 
measurements by Meyer et~al. (2007). The clustering in the modified version of the 
Bower et~al. model (MHIBow06) best agrees with the HIPASS results. 

One of the primary science goals of the Square Kilometre Array (SKA) is to make a high 
precision measurement of large-scale structure in the galaxy distribution. By measuring 
the apparent size of baryonic acoustic oscillations (BAO) at a particular redshift, the 
cosmological distance to that redshift can be derived, thereby constraining the equation 
of state of the dark energy. By combining the galaxy formation model with a very large 
volume N-body simulation ($ 1 h^{-3} {\rm Gpc}^{3}$), we have been able to demonstrate 
that galaxy samples constructed on the basis of cold gas mass can trace the BAO with the 
same fidelity as an near-infrared selected sample with the same number density of galaxies. 

The key remaining question is how effectively do HI and optical redshift surveys sample 
the available geometrical volume and how does this translate into an error on the dark energy 
equation of state parameter? The effective survey volume varies substantially between 
HI surveys of different duration and for different assumptions about the split between 
atomic and molecular hydrogen. However, at least for the case of a cosmological constant, 
these differences occur in a redshift range which has little impact on the derived error 
on the equation of state. We find that HI surveys are comparable to the most ambitious 
near-infrared spectroscopic surveys currently under discussion, and will give a factor of 
two smaller error on $w$ than a slitless H-$\alpha$ redshift survey; all are bone fide Stage 
V experiments in the Dark Energy Task Force nomenclature (Albrecht et~al. 2006). 
The uncertainty in the ratio of molecular to atomic hydrogen is one of the major uncertainties 
at present, and leads to larger differences in the predicted counts of HI emitters than the 
choice of galaxy formation model. The fraction of molecular hydrogen is thought to depend upon the local 
conditions in the interstellar medium. This question requires further modelling 
(e.g. Krumholz, McKee \& Tumlinson 2009), augmented by observations of the HI and CO 
distribution in nearby galaxies, for example by HI surveys on the SKA pathfinder MeerKAT 
and CO measurements using the Atacama Large Millimeter/submillimeter Array (Wootten 2008).

\section*{Acknowledgments}
HSK acknowledges support from the Korean Government's 
Overseas Scholarship. 
CSF acknowledges a Royal Society Wolfson Research Merit Award. 
This work was supported in part by grants from the Science and Technology 
Facilities Council at Durham and Leicester (CP).

\end{document}